\documentclass{article}

\usepackage{arxiv}

\usepackage[utf8]{inputenc} 
\usepackage[T1]{fontenc}    
\usepackage[numbers]{natbib}
\usepackage{hyperref}       
\usepackage{url}            
\usepackage{booktabs}       
\usepackage{amsfonts}       
\usepackage{nicefrac}       
\usepackage{microtype}      
\usepackage{lipsum}
\usepackage{fancyhdr}       
\usepackage{graphicx}       
\graphicspath{{media/}}     

\usepackage[title]{appendix}
\usepackage{amsmath} 
\usepackage{amssymb}
\usepackage{bm}
\usepackage{graphicx}
\usepackage{float}
\usepackage{subcaption}
\usepackage{xcolor}
\usepackage{mathrsfs}
\usepackage{mathptmx}
\usepackage{multirow}
\usepackage[ruled,vlined,linesnumbered]{algorithm2e}

\definecolor{edits}{rgb}{0, 0, 0}
\definecolor{oct15}{rgb}{0 0,0}

\usepackage[section]{placeins}
  
\title{Robust Bayesian state and parameter estimation framework for stochastic dynamical systems with combined time-varying and time-invariant parameters
}

\author{
  Philippe Bisaillon \\
  Department of Civil and Environmental Engineering\\
  Carleton University \\
  Ottawa, ON, Canada \\
   \And
  Brandon Robinson\\
  Department of Civil and Environmental Engineering\\
  Carleton University \\
  Ottawa, ON, Canada\\
   \And
  Mohammad Khalil
\thanks{\textit{Sandia National Laboratories is a multimission laboratory managed and operated by National Technology and Engineering Solutions of Sandia, LLC., a wholly owned subsidiary of Honeywell International, Inc., for the U.S. Department of Energy’s National Nuclear Security Administration under contract DE-NA-0003525.}} \\ 
  Quantitative Modeling \& Analysis Department\\
  Sandia National Laboratories \\
  Livermore, CA, United States\\
   \And
  Chris L. Pettit\\
  Aerospace Engineering Department \\
  US Naval Academy\\
  Annapolis, MD, United States\\
   \And
  Dominique Poirel \\
  Department of Mechanical and Aerospace Engineering\\
  Royal Military College of Canada \\
  Kingston, ON, Canada \\
   \And
  Abhijit Sarkar \\
  Department of Civil and Environmental Engineering\\
  Carleton University \\
  Ottawa, ON, Canada\\
}

\begin{document}
\maketitle

\begin{abstract}
We consider state and parameter estimation for a dynamical system having both time-varying and time-invariant parameters. It has been shown that the robustness of the Markov Chain Monte Carlo (MCMC) algorithm for estimating time-invariant parameters alongside nonlinear filters for state estimation provided more reliable estimates than the estimates obtained solely using nonlinear filters for combined state and parameter estimation. In a similar fashion, we adopt the extended Kalman filter (EKF) for state estimation and the estimation of the time-varying system parameters, but reserve the task of estimating time-invariant parameters to the MCMC algorithm. In a standard method, we augment the state vector to include the original states of the system and the subset of the parameters that are time-varying. Each time-varying parameter is perturbed by a white noise process, and we treat the strength of this artificial noise as an additional time-invariant parameter to be estimated by MCMC, circumventing the need for manual tuning. Conventionally, both time-varying and time-invariant parameters are appended in the state vector, and thus for the purpose of estimation, both are free to vary in time. However, allowing time-invariant system parameters to vary in time introduces artificial dynamics into the system, which we avoid by treating these time-invariant parameters as static and estimating them using MCMC. Furthermore, by estimating the time-invariant parameters by MCMC, the augmented state is smaller and the nonlinearity in the ensuing state space model will tend to be weaker than in the conventional approach. We illustrate the above-described approach for a simple dynamical system in which some model parameters are time-varying, while the remaining parameters are time-invariant. 
\end{abstract}


\section{Introduction}

The focus of this paper is to present a robust algorithm for the combined state and parameter estimation of nonlinear dynamical models having both time-varying and time-invariant parameters. It is an extension of the methods described by Bisaillon et al. \cite{bisaillon2015bayesian} which previously considered combined state and parameter estimation using a combination of nonlinear filters and Markov Chain Monte Carlo (MCMC) methods for systems having time-invariant parameters. Furthemore, this paper builds on the conclusions by Khalil et al. \cite{khalil2015estimation}, who investigated the use of nonlinear filters, namely, the ensemble Kalman filter (EnKF) and the particle filter (PF), for combined state and parameter estimation for systems having time-invariant parameters, and compared their performance to a benchmark consisting of a nested framework that leveraged MCMC for parameter estimation and the ensemble Kalman filter for state estimation alone. \textcolor{edits}{The use of parameter adaptive filtering for  the estimation involves the augmentation of the state space to include the model parameters in addition to the system states \cite{chui2017kalman,stengel1994optimal}}. The non-linearity in the augmented state is typically stronger than in the underlying dynamical system itself, and thus filters that may be well-suited for state estimation, may perform inadequately when considering the estimates of time-invariant parameters. Hence, we resort to employing the nonlinear filters exclusively for the estimation of the system states and parameters that are known to vary in time, while relying on MCMC for the estimation of time invariant parameters. 

\textcolor{edits}{A point of difficulty that arises from the augmentation of the state space is in determining the model error covariance associated with the parameters. This covariance matrix can be updated recursively, or can be static. It is important to note that further augmentation of the state vector cannot reliably be used to directly estimate the model error covariance matrix \cite{delsole2010state, stroud2007sequential}. In some applications, it may be necessary to jointly estimate the model error and measurement error covariances \cite{tandeo2020review}, however we assume the statistics of the measurement errors can be determined from the properties of the sensor devices. In such cases, it is only necessary to estimate the covariance of the model error.}

\textcolor{edits}{In similar applications to the current paper, relevant to damage detection in civil engineering structures (i.e. detecting changes in structural stiffness), a combination of the EKF and Robbins-Monro method \cite{robbins1951stochastic} have been used to recursively estimate the process noise (model error) covariance of the augmented state space \cite{yang2020structure,amini2021online}. Some alternative approaches used in the literature include adaptive tracking techniques which involve recursively optimizing the step size between sequential parameter estimates \cite{yang2006adaptive} and defining a parameterized model error covariance, and subsequently estimating these parameters \cite{dee1995line,storvik2002particle}. As noted in \cite{andrieu2003online}, these filtering-based approaches often result in unrealistically precise estimates of the parameters after a small number of iterations, effectively resulting in point estimates of the parameters as their posterior distributions become Dirac delta functions. In the case where the covariance matrix is static, rather than recursively estimated, some degree of uncertainty in the parameters is preserved. In such cases, the parameters of the covariance matrix can be tuned manually, however, this method is not robust and can lead to sub-optimal results.} The use of MCMC for estimating the time-invariant parameters, including the parameters of the model error covariance matrix, provides many benefits, which offset the additional computational burden it introduces. In cases where the majority of the system parameters are time-invariant, the proposed framework reduces the dimensionality of the augmented state space that would otherwise be required. This also has the added benefit of reducing the degree of nonlinearity in the system that would otherwise have been introduced by appending static parameters to the state vector. Finally, it removes artificial dynamics that are introduced when modelling parameters that are known to be time-invariant as time-varying parameters, as is necessary when performing joint state and parameter estimation using filtering methods.

The Kalman filter is the optimal estimator for the states of a linear dynamical systems with measurements having additive Gaussian noise \cite{kalman1960new}. In the numerical example in this paper, the underlying dynamical system is linear in nature, however, when we append the time-varying parameters to the state vector, the state space model becomes nonlinear, requiring the use of nonlinear filters. In this case, we use a nonlinear extension of the Kalman filter, the Extended Kalman filter (EKF), which relies on the local linearization the model and measurement operators. The EKF is suitable in either weekly nonlinear systems or for applications having dense data, as it is based on the assumption that the elements of the state vector have approximately Gaussian distributions. In the current investigation, EKF was found to be adequate (due to dense observations, whereby the state remains nearly Gaussian). However, for highly non-Gaussian models, more robust Monte Carlo (MC)-based filters (e.g. EnKF or PF) will be required for the evaluation of the likelihood function within the MCMC sampling \cite{khalil2009nonlinear,khalil2015estimation,bisaillon2015bayesian}.

Section \ref{problemstatement} provides the requisite mathematical details of the aforementioned framework. Relevant background information is provided for the use of nonlinear filters for the recursive estimation of the system states and time-varying parameters and the associated calculation of the likelihood function. The implementational details for the Transitional MCMC (TMCMC) algorithm is then described as necessary for the estimation of time-invariant parameters. Section \ref{results} presents a case study of a linear single-degree-of-freedom (SDOF) oscillator that is driven by a stochastic forcing term, and experiences structural degradation, represented by a change in stiffness. Synthetic data which represent the observations of the displacement of the oscillator are used for calibration. We propose a set of candidate models and perform Bayesian model comparison to assess the abilities of the respective models in capturing the observed behaviour by estimating the combination of time-varying and time-invariant parameters. We consider each of these models in three different scenarios. In Section \ref{baseline}, we consider a small, but sudden change in stiffness. In Section \ref{sudden} we consider the case where the sudden change in stiffness is much more significant. Finally, in Section \ref{slowsnap} we consider the case where there is a gradual change in stiffness, represented by a linear decrease over the duration of the observations. In the first scenario, the effect of the change in stiffness on the observed dynamics is difficult to detect visually, which is designed to assess the ability of the proposed model in detecting subtle changes in system parameters within the proposed framework. In the second case, the observed dynamics differ significantly before and after the sudden change in stiffness, which can be easily identified visually, but will test the proposed models' ability to adapt to a significant sudden change in the system parameters. The third scenario seeks to evaluate each model's ability to generalize to a different form of structural degradation. We compare the performance of all the proposed models in each of the above described scenarios in Sections \ref{c1_modelsel}, \ref{c2_modelsel} and \ref{c3_modelsel}, using an evidence-based Bayesian model selection approach \cite{bisaillon2015bayesian,sandhu2014bayesian,sandhu2016bayesian,sandhu2017bayesian}.

\textcolor{edits}{In this paper, we demonstrate the benefit of estimating the time-invariant model error strength from the original stochastic dynamical system, as well as the artificial noise strength that is introduced through the process of augmenting the state vector to include the estimation of static parameters.}

\section{Problem Statement}\label{problemstatement}

This section contains a description of the Bayesian algorithm for parameter estimation. Consider a model described by \cite{bisaillon2015bayesian,khalil2015estimation,khalil2009nonlinear,jazwinski2007stochastic,evensen2009data}

\begin{equation}
\mathbf{u}_{k+1} = \mathbf{g}_k(\mathbf{u}_k, \bm{\Phi}_t,  \bm{\Phi}_s, \mathbf{q}_k)
\end{equation}

where we denote the model operator $\mathbf{g}_k$, system states $\mathbf{u}_k$, model error $\mathbf{q}_k$ with a prescribed joint probability density function (pdf), $\bm{\Phi}_s$ denotes the time-invariant parameters, and $\bm{\Phi}_t =\left\lbrace  \bm{\Phi}_{t(0)},\hdots, \bm{\Phi}_{t(k)} \right\rbrace$ is the vector of time-varying parameters at discrete instances of time along the computational grid. 
All of the parameters are to be estimated using sensor data.  Sensor data relate to the state through the following measurement equation \cite{bisaillon2015bayesian,khalil2015estimation,khalil2009nonlinear,jazwinski2007stochastic,evensen2009data}

\begin{equation}
 \mathbf{d}_{j} = \mathbf{h}_{j} (\mathbf{u}_{d(j)}, \mathbf{\varepsilon}_{j}) \label{eq:meas}
\end{equation}

where $\mathbf{d}_{j}$ is the observation vector at the time step $t_{d(j)}$ and $\mathbf{\varepsilon}_j$ is the measurement noise modelled as a zero-mean random vector having a known covariance matrix.

If we consider an augmented state $\mathbf{x}_k = \{\mathbf{u}_k, \bm{\Phi}_{t(k)}\}$, the joint posterior of the system state, time-invariant and time-varying parameters of model $\mathcal{M}$, conditioned on data $\mathbf{D} = \{\mathbf{d}_0, \hdots,\mathbf{d}_J\}$ can be represented as follows \cite{bisaillon2015bayesian,khalil2015estimation,khalil2009nonlinear}

\begin{equation}
\text{p}(\mathbf{u}_1,...,\mathbf{u}_k,\bm{\Phi}_{t(0)},\hdots,\bm{\Phi}_{t(k)}, \bm{\Phi}_s | \mathbf{D}, \mathcal{M})  = \text{p}(\mathbf{x}_1,...,\mathbf{x}_k, \bm{\Phi}_s | \mathbf{D}, \mathcal{M}) = \text{p}(\mathbf{x}_1,...,\mathbf{x}_k | \mathbf{D}, \bm{\Phi}_s, \mathcal{M}) \text{p}(\bm{\Phi}_s | \mathbf{D},\mathcal{M}) \\
\end{equation}

For parameter estimation we are interested in the posterior distribution of our parameters,  $\text{p}(\bm{\Phi}_{t(1)},...,\bm{\Phi}_{t(k)}, \bm{\Phi}_s | \mathbf{D}, \mathcal{M})$. We employ the Transitional Markov Chain Monte Carlo (TMCMC) method \cite{ching2007transitional} for the estimation of static parameters $\bm{\Phi}_s$, while we employ the extended Kalman filter (EKF) for the concurrent estimation of the state $\textbf{u}$ and time-varying parameter vector $\bm{\Phi}_t$. The state estimation algorithm is outlined in Section \ref{section:stateestimation}.  Thus the state estimation gives us the posterior of our augmented state 

\begin{equation}
\text{p}(\mathbf{u}_1,...,\mathbf{u}_k,\bm{\Phi}_{t(1)},...,\bm{\Phi}_{t(k)} | \bm{\Phi}_s , \mathbf{D}, \mathcal{M}) = \text{p}(\mathbf{x}_1,...,\mathbf{x}_k | \bm{\Phi}_s , \mathbf{D}, \mathcal{M}) 
\end{equation}

\subsection{Framework for estimating time-varying and time-invariant parameters}

Consider the model $\mathcal{M}$ having the unknown parameters $\bm{\Phi}_s$, the estimated values of the parameter after assimilating measurements $\mathbf{D}$ is
\begin{equation}
\text{p}(\bm{\Phi}_s | \mathbf{D}, \mathcal{M}) = \frac{\text{p}(\mathbf{D} | \bm{\Phi}_s, \mathcal{M}) \text{p}( \bm{\Phi}_s | \mathcal{M})}{\text{p}(\mathbf{D} | \mathcal{M})} \propto \text{p}(\mathbf{D} | \bm{\Phi}_s \mathcal{M}) \text{p}( \bm{\Phi}_s | \mathcal{M}) ,\label{eq:bayesmodel}
\end{equation}
where $\text{p}(\bm{\Phi} |\mathbf{D},\mathcal{M}_i)$ is the parameter posterior, $ \text{p}(\bm{\Phi}|\mathcal{M}_i)$ is the parameter prior distribution, and $\text{p}(\mathbf{D} |\bm{\Phi},\mathcal{M}_i)$ is the likelihood function. The likelihood function is evaluated as the product of the likelihood computed at each data point ~\cite{bisaillon2015bayesian,bisaillon2022combined}

\begin{equation}
\text{p}(\mathbf{D} | \bm{\Phi}_s, \mathcal{M}) = \prod_{j=1}^J \text{p}(\mathbf{d}_j | \bm{\Phi}_s, \mathcal{M}), \label{eq:likelihood}
\end{equation}
with

\begin{align}
\text{p}(\mathbf{d}_j | \bm{\Phi}_s, \mathcal{M}) = \int_{-\infty}^{\ \infty} \text{p}( \mathbf{x}_{d(j)} | \mathbf{x}_{d(j)-1}, \bm{\Phi}_s) \text{p}(\mathbf{d}_j | \mathbf{x}_{d(j)} , \bm{\Phi}_s) \text{d} \mathbf{x}_{d(j)}.\label{eq:lik}
\end{align}

The computation of the likelihood function in Eq.~\eqref{eq:lik} involves a state estimation task through nonlinear filters (e.g. \cite{bisaillon2015bayesian,khalil2015estimation,khalil2009nonlinear,chen2003bayesian}), as outlined in Section~\ref{section:stateestimation}. 

Since the term $\text{p}(\mathbf{D} | \mathcal{M})$ in the denominator of Eq. (\ref{eq:bayesmodel}) is not readily known and is analytically intractable in general, the posterior distribution $\text{p}(\bm{\Phi}_s | \mathbf{D}, \mathcal{M})$ is only known up to a constant of proportionality. To sample from the posterior distribution, TMCMC is used as described in Section~\ref{section:tmcmc}. An advantage of TMCMC over other MCMC algorithms for the current application is that samples are independent and TMCMC is well suited to handle multimodal posterior pdfs \cite{ching2007transitional}. 

\subsubsection{State Estimation}\label{section:stateestimation}
In this application, the state estimation procedure serves two purposes. The first is to estimate the state of the system $\mathbf{u}$ given the proposed model, the set of static parameters, and the incoming noisy sensor data. The second is to estimate the time-varying parameters through the augmented state vector $\mathbf{x}$. In the first case if the state-space model is linear, we can use the Kalman filter (KF) for state estimation. As we augment the state to include the time varying parameters, we have now introduced nonlinearity into the system, hence nonlinear filters are required. Here we use the extended Kalman filter (EKF) for all cases, noting that this imposes the assumption that the state pdfs are approximately Gaussian, which generally holds for weakly nonlinear systems. 

The EKF imposes a Gaussian assumption on the state such that $\mathbf{x}_k \sim \mathcal{N}(\mathbf{x}_k^a,\mathbf{P}_k^a)$, denoting a Gaussian pdf with mean $\mathbf{x}_k^a$ and covariance $\mathbf{P}_k^a$. Similarly, the model error (process noise) is assumed to be normally distributed with $\mathbf{q}_k \sim \mathcal{N}(\mathbf{0},\mathbf{Q}_k)$. For points along the computational grid, indexed by $k$, that do not coincide with a data point, indexed by $d(j)$, the current mean state estimate and its uncertainty are forecasted as follows 

\begin{align}
\mathbf{x}_{k+1}^f&=\mathbf{g}_k(\mathbf{x}_k^a,\mathbf{f}_k, \mathbf{0}), \\
\mathbf{P}_{k+1}^f&= \mathbf{A}_k\mathbf{P}_k^a\mathbf{A}_k^T + \mathbf{B}_k\mathbf{Q}_k\mathbf{B}_k^.
\end{align}
where the Jacobian matrices

\begin{align}
\mathbf{A}_k &= \left. \frac{\partial \mathbf{g}_{k}(\mathbf{x}_k, \bm{\varepsilon}_k)}{\partial \mathbf{x}_{k}} \right\vert_{\mathbf{x}_k = \mathbf{x}_{k}^f, \mathbf{\varepsilon}_k = \mathbf{0}} \label{eq:ekf_update1}, \\
\mathbf{B}_k &=  \left. \frac{\partial \mathbf{g}_{k}(\mathbf{x}_{k}, \bm{\varepsilon}_k)}{\partial \bm{\varepsilon}_k} \right\vert_{\mathbf{x}_k = \mathbf{x}_k^f, \mathbf{\varepsilon}_k = \mathbf{0}}  \label{eq:ekf_update2}.
\end{align}

When the numerical integration grid coincides with a measurement, the analysis step is performed, allowing for the mean and covariance of the state estimate to be updated. The measurement operator is given in Eq. (\ref{eq:meas}), where the measurement noise is assumed to be Gaussian $\varepsilon_j \sim \mathcal{N}(\mathbf{0},\bm{\Gamma}_j)$ resulting in the following update step (e.g., \cite{bisaillon2015bayesian,khalil2015estimation,khalil2009nonlinear,sandhu2014bayesian,sandhu2016bayesian,jazwinski2007stochastic,evensen2009data}):

\begin{align}
\mathbf{x}_{d(j)}^{a}&=\mathbf{x}_{d(j)}^f + \mathbf{K}_{d(j)}(\mathbf{d}_j - \mathbf{h}_{d(j)}(\mathbf{x}_{d(j)}^f,\mathbf{0})). \\
\mathbf{P}_{d(j)}^a&= (\mathbf{I} - \mathbf{K}_j\mathbf{C}_j)\mathbf{P}_{d(j)}^f.\\
\mathbf{K}_{d(j)}^a&= \mathbf{P}_{d(j)}^f \mathbf{C}_j^T \left[\mathbf{C}_j \mathbf{P}_{d(j)}^f \mathbf{C}_j^T + \mathbf{D}_j \bm{\Gamma}_{j} \mathbf{D}_j^T  \right]^{-1},
\end{align}
where the Jacobian matrices $\mathbf{C}_j $ and $\mathbf{D}_j$ are evaluated as

\begin{align}
\mathbf{C}_j &= \left. \frac{\partial \mathbf{h}_{d(j)}(\mathbf{x}_{d(j)}, \bm{\varepsilon}_j)}{\partial \mathbf{x}_{d(j)}} \right\vert_{\mathbf{x}_{d(j)} = \mathbf{x}_{d(j)}^f, \mathbf{\varepsilon}_{j} = \mathbf{0}} \label{eq:ekf_update1}, \\
\mathbf{D}_j &=  \left. \frac{\partial \mathbf{h}_{d(j)}(\mathbf{x}_{d(j)}, \bm{\varepsilon}_j)}{\partial \bm{\varepsilon}_j} \right\vert_{\mathbf{x}_{d(j)} = \mathbf{x}_{d(j)}^f, \mathbf{\varepsilon}_j = \mathbf{0}}  \label{eq:ekf_update2}.
\end{align}
Note that for linear model operators and measurements with additive Gaussian noise, the above described EKF reduces to the simple KF, as the Jacobian matrices $\mathbf{A}_j$, $\mathbf{B}_j$, $\mathbf{C}_j$, and $\mathbf{D}_j$ become identity matrices.

Critically, when data are available, we can also compute the likelihood function in Eq. (\ref{eq:lik}) according to \cite{bisaillon2015bayesian,khalil2015estimation,khalil2009nonlinear,sandhu2014bayesian,sandhu2016bayesian,khalil2013bayesian}

\begin{align}
 \text{p}(\mathbf{d}_j\vert \bm{\Phi} ,\mathcal{M}_i) =  \mathcal{N}\left(\mathbf{d}_j | \mathbf{h}_{j}\left(\mathbf{x}_{d(j)}^f, \mathbf{0}\right), \bm{\Sigma '}\right)  \label{eq:int_ekf}
\end{align}
where 
\begin{equation}
\bm{\Sigma '} = \mathbf{C}_j \mathbf{P}_{d(j)}^f \mathbf{C}_j^T + \mathbf{D}_j \bm{\Gamma}_j \mathbf{D}_j^T
\end{equation}

\subsubsection{Transitional Markov Chain Monte Carlo (TMCMC)}\label{section:tmcmc}
In the description below, we omit the subscript from $\bm{\Phi}_s$ to avoid confusing the  $s$ (which denotes the set of time-invariant parameters) with the indices used in the TMCMC procedure. Hence, we simply use the symbol $\bm{\Phi}$ to denote the time-invariant parameters to be estimated by the TMCMC algorithm, while the above described state estimation procedure accounts for the time-varying parameters. 

TMCMC begins by sampling from the prior distribution followed by sampling from a series of consecutive intermediate distributions which converge towards the posterior distribution \cite{ching2007transitional}. Based on adaptive Metropolis-Hastings MCMC, the methodology draws inspiration from the concept of simulated annealing \cite{beck2002bayesian}. Consider the distribution \cite{ching2007transitional}
\begin{equation}
\text{p}_j( \bm{\Phi} ) \propto \text{p}( \bm{\Phi} | \mathcal{M} ) \text{p}( \mathbf{D} | \bm{\Phi} , \mathcal{M} )^{p_j}
\end{equation}
where subscript $j = 0, 1, \dots, m$ denotes the stage number, and the exponent $p_j$ follows $p_0 =0 < p_1 < \dots < p_m = 1$. For $j=0$, sampling from the distribution $\text{p}_j( \bm{\Phi})$ reduces to sampling from the prior, and for $j = m$, the samples are drawn from the posterior distribution. Careful selection of the number of stages, $m$, and the exponents of the intermediate distributions for $j = 1, \hdots, m-1$ are critical; enough stages should be used to ensure that the intermediate distributions approach the unnormalized posterior without significant changes between stages, but not so many stages that convergence to the posterior becomes overly expensive. For efficient implementation of TMCMC, the exponent $p_{j+1}$ may be selected for the subsequent stage such that the coefficient of variation of the plausibility weights $\text{p}( \mathbf{D} | \bm{\Phi} , \mathcal{M} )^{p_{j+1}-p_j}$ does not exceed a specified threshold \cite{ching2007transitional}. In this study, we use a target coefficient of variation of 1 to compute a suitable value of $p_{j+1}$ using the bisection method \cite{bisaillon2022combined,sandhu2021nonlinear}. While more efficient methods could be implemented to reduce the number of iterations required for convergence, this step is not particularly expensive.

To initialize the algorithm in the zeroth stage (indexed by $j=0$), $N$ samples of $\bm{\Phi}$ are generated from the parameter prior ($\text{p}_0(\bm{\Phi}) = \text{p}(\bm{\Phi}\vert \mathcal{M})$) using Monte Carlo sampling. Then for all stages $j=0, \hdots, m-1$, the following steps apply. The value of $p_{j+1}$ is calculated according to the desired criteria and then the plausibility weights $w(\bm{\Phi}_{j,k})$ are calculated for each sample (indexed by $k = 1,\hdots, N$) as \cite{ching2007transitional}

\begin{equation}\label{eq:tmcmc_w}
w(\bm{\Phi}_{j,k})  = \frac{\text{p}( \bm{\Phi}_{j,k} \vert \mathcal{M} ) \text{p}( \mathbf{D} | \bm{\Phi}_{j,k} , \mathcal{M} )^{p_{j+1}} }{\text{p}( \bm{\Phi}_{j,k} \vert \mathcal{M} )\text{p}( \mathbf{D} | \bm{\Phi}_{j,k} , \mathcal{M} )^{p_j} }  = \text{p}( \mathbf{D} | \bm{\Phi}_{j,k} , \mathcal{M} )^{p_{j+1}-p_j}.
\end{equation}

For each sample, a single Metropolis-Hastings (MH) step is used. The new samples are proposed with
\begin{equation}
\bm{\Phi}_{j+1,k} \sim \mathcal{N}\left( \bm{\Phi}_{j,k}, \bm{\Sigma}_j \right)
\end{equation}
where $\mathcal{N}(\bm{\Phi}_{j,k}, \bm{\Sigma}_j )$ denotes a Gaussian random vector with mean vector $\bm{\Phi}_{j,k}$ and covariance matrix $\bm{\Sigma}_j$. The covariance matrix $\bm{\Sigma}_j$ is the product of the estimated covariance of $\text{p}_{j+1}(\bm{\Phi})$ and scaling factor $\beta^2$  \cite{ching2007transitional,roberts2001optimal}

\begin{equation}\label{eq:tmcmc_propcov}
\Sigma_j = \beta^2 \sum_{k=1}^{N_j} w(\bm{\Phi}_{j,k}) \left\{\bm{\Phi}_{j,k} - \bm{\mu}_j\right\}\left\{\bm{\Phi}_{j,k} - \bm{\mu}_j\right\}^T
\end{equation}
with

\begin{equation}
\bm{\mu}_j =\frac{\sum_{k=1}^{N_j} w(\bm{\Phi}_{j,k}) \bm{\Phi}_{j,k}}{\sum_{k=1}^{N_j} w(\bm{\Phi}_{j,k}) }
\end{equation}

The samples $\bm{\Phi}_{j+1,k} $ are accepted with probability $\alpha$ 

\begin{equation}
\alpha = \min\left(1.0, \frac{\text{p}( \bm{\Phi}_{j+1,k} | \mathcal{M} ) \text{p}( \mathbf{D} | \bm{\Phi}_{j+1,k} , \mathcal{M} )^{p_i}}{\text{p}( \bm{\Phi}_{j,k} | \mathcal{M} ) \text{p}( \mathbf{D} | \bm{\Phi}_{j,k} , \mathcal{M} )^{p_i}}\right)
\end{equation}

\begin{algorithm}[H]
\SetAlgoLined
Define the number of samples to be generated per stage $N$\;
Initialize $j = 0$\;
Initialize $p_0 = 0$ \;
Generate samples $\{\bm{\Phi}_{0,k}; \ k= 1,\hdots, N\}$ from the prior pdf of the static parameters $\text{p}( \bm{\Phi} | \mathcal{M})$\;
 \While{$p_j < 1$}{
Compute the likelihood $\text{p}( \mathbf{D} | \bm{\Phi}_{j,k} , \mathcal{M} )$ for each sample in the $j^{th}$ stage according to Eq. (\ref{eq:lik}) and Eq. (\ref{eq:int_ekf})\;
Select $p_{j+1}$ s.t. the coefficient of variation of $\text{p}( \mathbf{D} | \bm{\Phi}_{j,k} , \mathcal{M} )^{p_{j+1}-p_j} = 1 $\;
Compute the weights of each sample according to Eq. (\ref{eq:tmcmc_w})\;
Compute the covariance of the proposal distribution according to Eq. (\ref{eq:tmcmc_propcov})\;
For each sample, perform a MH stage\;
$j = j+1$ \;
 }
Generate $N$ samples from the unnormalized posterior pdf of the static parameters $\text{p}( \bm{\Phi} | \mathcal{M} ) \text{p}( \mathbf{D} | \bm{\Phi} , \mathcal{M} )$\;
 \caption{TMCMC algorithm}\label{tmcmc_algo}
\end{algorithm}

Note that in an alternative approach, if the modeller wishes to manually define the number of stages $m$, a schedule for $p_j$ should be defined, replacing line 7 in Algorithm \ref{tmcmc_algo}
 
\subsection{Model Selection}\label{section:modelsel}
The current paper seeks to illustrate the benefits of using a combination of nonlinear filters and MCMC for calibrating systems having both time-varying and time-invariant parameters. To do so, for each of the three test cases we consider in Section \ref{results}, we perform Bayesian model selection to quantify the performance of models that are calibrated using the proposed approach. The basis upon which we quantify the relative performance of each of the candidate models is based on the posterior probability of model $\mathcal{M}_i$ \cite{bisaillon2015bayesian,sandhu2014bayesian}

\begin{equation}
\text{p}(\mathcal{M}_i \vert \mathbf{D}, \bm{\mathcal{M}}) = \frac{\text{p}(\mathbf{D}\vert \mathcal{M}_i) \text{p}(\mathcal{M}_i \vert  \bm{\mathcal{M}})}{\sum_{j=i}^{N_M} \text{p}(\mathbf{D}\vert \mathcal{M}_j) \text{p}(\mathcal{M}_j \vert  \bm{\mathcal{M}})}.\label{eq:model_posterior}
\end{equation}

In the case where all model priors $\text{p}(\mathcal{M}_i \vert  \bm{\mathcal{M}})$ are equal, Eq. (\ref{eq:model_posterior}) reduces to an expression for the normalized model evidence. The model evidence, or more precisely, the log Evidence, implicitly balances the trade-off between the average data-fit and model complexity, leading to the so-called  quantitative Occam's razor  \cite{bisaillon2022combined,sandhu2014bayesian,muto2008bayesian},

\begin{equation}
\underbrace{\ln \text{p}(\mathbf{D}\vert \mathcal{M})}_{\text{Log evidence}} = \underbrace{\mathbb{E}\left[\ln \text{p}(\mathbf{D} \vert \bm{\Phi}, \mathcal{M})\right]}_{\text{Average goodness-of-fit}} -  \underbrace{\mathbb{E}\left[\ln \frac{\text{p}(\bm{\Phi}\vert\mathbf{D}, \mathcal{M})}{\text{p}(\bm{\Phi} \vert \mathcal{M}) } \right]}_{\text{Information gain/Complexity}} , \label{eq:occam}
\end{equation}

where the expectation is with respect to the posterior of the model parameter $\bm{\Phi}$. While it is not used here, it is worth noting that the TMCMC algorithm provides a convenient estimate for the model evidence $\text{p}(\textbf{D}\vert \mathcal{M})$, \cite{ching2007transitional}. In this study we make use of the Chib-Jealiazkov method for calculation of the model evidence \cite{bisaillon2015bayesian,sandhu2014bayesian,bisaillon2022combined,chib2001marginal}

\section{Numerical Results}\label{results}

\begin{figure}[!htb]
\centering
\includegraphics[width=\textwidth]{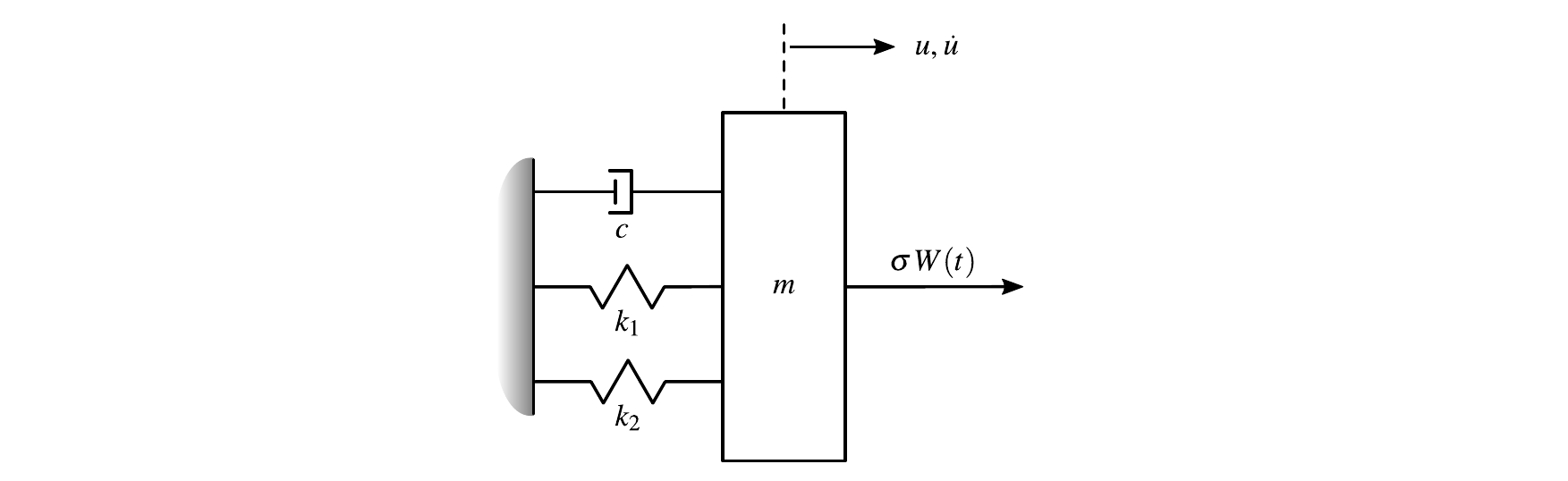}
\caption{Mass spring damper system with random forcing}
\label{fig:mck1}
\end{figure}

Consider a damped SDOF harmonic oscillator with two parallel springs driven by a random forcing as in Figure \ref{fig:mck1}, having the following equation of motion

\begin{equation}\label{eq:mck_dynamics}
m \ddot{u} + c \dot{u} + K(t)u = \sigma W(t) .
\end{equation}
where the overall stiffness of the system $K(t)$ is given by the sum of two parallel springs, having stiffness $k_1$ and $k_2$, respectively.

The spring having stiffness $k_2$ is allowed to degrade, either experiencing a sudden failure (as in Case 1 and Case 2), modelled as a piece-wise step function occurring at time $t_s$,

\begin{equation}
K(t) = \begin{cases} 
      k_1 + k_2 &\quad \text{if } t < ts \\
       k_1 &\quad \text{if } t \geq t_s \\
     \end{cases},\label{eq:gen12}
\end{equation}
or as a linear degredation in time (as in Case 3), 
\begin{equation}
K(t) = k_1 + k_2\left(1-\frac{t}{20}\right). \label{eq:gen3}
\end{equation}
While the overall system stiffness changes in time due to this structural degradation, the stiffness coefficients $k_1$ and $k_2$ themselves, as well as the mass $m$, damping coefficient $c$, and the magnitude of the random forcing $\sigma$ are time-invariant.

Using the Euler-Maruyama scheme \cite{kloeden1992stochastic,khalil2015estimation,bisaillon2022combined}, the discrete state-space representation of this system, where $\mathbf{x} = \{u,\dot{u}\}^T$, is given by

\begin{subequations}
\begin{align}
\mathbf{x}_{1,k+1} &= \mathbf{x}_{1,k} + \Delta t \mathbf{x}_{2,k}, \\
\mathbf{x}_{2,k+1} &= (-\Delta t \ k /m) \mathbf{x}_{1,k} + (1 - \Delta t \ c/m) \mathbf{x}_{2,k} + \sqrt{\Delta t} \sigma \ \xi_{k} \label{eq:model3c}
\end{align}
\end{subequations}

where $K_k = K(t_k)$, and $\xi_k$ is drawn from a Gaussian distribution with zero mean and unit variance. Noisy measurements $d_k$  $(k = 1,\hdots, K) $ of the displacement $u_k$ are available at discrete time instances according to 
\begin{equation}
d_k = u_k + \varepsilon_k, 
\end{equation}
where $\varepsilon_k$ is a zero mean Gaussian random variable with a standard deviation of 10 mm. A 20 second realization of this system is generated according to the parameters and initial conditions summarized in Table \ref{tab:genmodel}. This signal is sampled at equal intervals (sampling rate of 25 Hz) and corrupted by additive Gaussian noise to generate the synthetic data. For reference, the root-mean-square of the displacement (pre-snap) was 34.6 mm, giving a noise-to-signal ratio of approximately 8.35\%.

\begin{table}[!htb]
\begin{center}
\begin{tabular}{llcll}
\hline
Parameter name & Symbol & Parameter value & Test Case & Units \\
\hline
\multirow{3}{*}{Undamaged spring stiffness} & \multirow{3}{*}{$k_1$} & 70 & Case 1 (\ref{baseline}) & [N/mm]\\
& & 10 &Case 2 (\ref{sudden}) & [N/mm]\\
&  & 70 & Case 3 (\ref{slowsnap}) &[N/mm]\medskip \\
\multirow{3}{*}{Damaged spring stiffness} & \multirow{3}{*}{$k_2$} & 10 &Case 1 (\ref{baseline}) & [N/mm]\\
& & 70 &Case 2 (\ref{sudden}) & [N/mm]\\
&  & 10 & Case 3 (\ref{slowsnap}) &[N/mm]\\
\hline
Mass & $m$ & 1 &All cases& [kg] \\
Damping coefficient & $c$ & 0.1 & All cases & [Ns/mm] \\
Magnitude of random forcing & $\sigma$ & 50  &All cases& [N] \\
Initial displacement & $u(0)$ & 50 &All cases&  [mm] \\
Initial velocity & $\dot{u}(0)$ & 0 & All cases& [mm/s] \\
\hline \hline
\end{tabular}
\caption{Parameters used in the data generation for the three test cases}\label{tab:genmodel}
\end{center}
\end{table}

A candidate set consisting of six distinct model structures are then defined for the inverse problem. Each candidate has the main underlying structure described by Eq. (\ref{eq:candidate0}),

\begin{equation}
m \ddot{u} + c \dot{u} + K(t)u = f(t), \label{eq:candidate0}
\end{equation}

where we consider the following six combinations of models for the time-varying stiffness parameter $K(t)$ and random forcing $f(t)$,

\begin{align}
\Big. \mathcal{M}_1:& \quad K(t) = \begin{cases} \stepcounter{equation}\tag{{\theequation}a} \label{eq:candidate1}
       k_1 + k_2 &\quad \text{if } t < t_s \\
       k_1 &\quad \text{if } t \geq t_s \\
     \end{cases} ,& f(t) &= \sigma W(t), \\
\Big. \mathcal{M}_2:& \quad K(t) = k, \tag{{\theequation}b} \label{eq:candidate2} & f(t) &= \sigma W(t), \nonumber \\
\Big. \mathcal{M}_3:& \quad K(t) = k,  \tag{{\theequation}c} \label{eq:candidate3} 	&\quad  \dot{f}(t) + \frac{1}{\tau} f(t) &= \sigma W(t), \nonumber\\
\Big. \mathcal{M}_4^a:& \quad \dot{K}(t) = \gamma_a q(t), \tag{{\theequation}d}\label{eq:candidate4a} &\quad f(t) &= \sigma W(t), \nonumber \\
\Big. \mathcal{M}_4^b:& \quad \dot{K}(t) = \gamma_b q(t), \tag{{\theequation}e}\label{eq:candidate4b} &\quad f(t) &= \sigma W(t), \nonumber \\
\Big. \mathcal{M}_5:& \quad \dot{K}(t) = \gamma q(t), \tag{{\theequation}f} \label{eq:candidate5}	&\quad f(t) &= \sigma W(t), \nonumber \\
\Big. \mathcal{M}_6:& \quad  \dot{K}(t) = \gamma q(t), \tag{{\theequation}g} \label{eq:candidate6} &\quad f(t) &= \sigma W(t). \nonumber 
\end{align}

Note that model $\mathcal{M}_1$ allows for a change in stiffness to be captured by a set of time-invariant parameters, but specifically models the change in stiffness as a step function. This model is therefore well-suited for capturing a sudden structural failure (as in Case 1 and 2) as it has the same form as the system that was used to generate the synthetic data, but it may not generalize well to the gradual degradation (as in Case 3). Model $\mathcal{M}_2$ consists of only time-invariant parameters and does not model any change in stiffness, thus providing a basis for assessing the benefits of trying to capture a change in stiffness explicitly. Model $\mathcal{M}_3$ also does not model the change in stiffness, but introduces a coloured noise model for the stochastic forcing \cite{bisaillon2022combined}, to assess whether the unmodelled physics captured better when the noise process is correlated rather than a white noise process. This introduces the parameter $\tau$, the relaxation time for the correlated noise process in Eq. (\ref{eq:candidate3}), which will be estimated concurrently with the other system parameters by MCMC. Models $\mathcal{M}_4$, $\mathcal{M}_5$ and $\mathcal{M}_6$ model the stiffness as a time-varying parameter by introducing artificial dynamics via the random process $q(t)$ in Eq. (\ref{eq:candidate4a}) to Eq. (\ref{eq:candidate6}). First, we consider two cases of the model  $\mathcal{M}_4$ (denoted by superscripts a, and b), where the strength of this artificial dynamics in stiffness is given by a known value $\gamma_a = 1$ and $\gamma_b = 10$. In the case of $\mathcal{M}_5$, we then estimate the strength of the artificial process noise, $\gamma$, thus relaxing some of the restrictions of $\mathcal{M}_4$. For both $\mathcal{M}_4$ and $\mathcal{M}_5$, we consider the nominal case where the initial mean for the stiffness parameter is accurately known (80 N/mm), and a secondary case where there is an erroneous initial mean for the stiffness parameter of 60 N/mm. This highlights the benefits of the flexibility of estimating the parameter $\gamma$ by MCMC, rather than relying on manual tuning of this parameter. Finally, $\mathcal{M}_6$ introduces a final degree of complexity as it estimates the initial mean for the stiffness parameter ($\mathbb{E}[K(0)]$).

For all cases, we use the following parameter prior distributions $\text{p}(\bm{\Phi}\vert \mathcal{M})$ detailed in Table \ref{tab:priors}. We adopt the use of non-informative priors with broad support for all parameters. Note that for models $\mathcal{M}_4^a$,  $\mathcal{M}_4^b$, and $\mathcal{M}_5$, we assign two fixed values for the initial mean of the system stiffness (assigning both the correct value of 80 N/mm, and an incorrect value of 60 N/mm in brackets), and only assign a prior for model $\mathcal{M}_6$, where we introduce it as a static parameter to be estimated.
\begin{table}[!h]
\centering
\begin{tabular}{clllllll}
\hline
Parameter &	$\mathcal{M}_1$ & 	$\mathcal{M}_2$ &	$\mathcal{M}_3$ &	$\mathcal{M}_4^a$ &	$\mathcal{M}_4^b$ &	$\mathcal{M}_5$ &	$\mathcal{M}_6$ \\
\hline																		
$k_1$	&	$\mathcal{U}(0,1000)$	& -	& -	&	- &	-	&-	&	-\\
$k_2$   &	$\mathcal{U}(0,1000)$	& - & - &	- &	-	&-	&	-\\
$t_s$	&	$\mathcal{U}(0,20)$		& -	& -	&	- &	-	&-	&	-\\
$K$		&	 -			&			$\mathcal{U}(0,1000)$	&$\mathcal{U}(0,1000)$	&-	&	-	&	-	&	-\\
$\mathbb{E}[K(0)]$	& -	& - & -	&  80 (*60)	&  80 (*60)	&80 (*60) &	$\mathcal{U}(0,100)$\\
$c$			&	$\mathcal{U}(0,5)$					&	$\mathcal{U}(0,5)$		  	&	$\mathcal{U}(0,5)$			&	$\mathcal{U}(0,5)$			&	$\mathcal{U}(0,5)$			&	$\mathcal{U}(0,5)$			&	$\mathcal{U}(0,5)$	\\
$\sigma$		&	$\mathcal{U}(0,1000)$ 	& $\mathcal{U}(0,1000)$	&	$\mathcal{U}(0,1000)$	&	$\mathcal{U}(0,1000)$ 	& 	$\mathcal{U}(0,1000)$	& $\mathcal{U}(0,1000)$	&	$\mathcal{U}(0,1000)$  \\
$\gamma$		&	-				          						 	&	-	&	-		&	1	& 10 &	$\mathcal{U}(0,1000)$ 	&	 $\mathcal{U}(0,1000)$ \\
$\tau$ &	-	&	-	&	$\mathcal{U}(0,10)$		&	-  	&	-	&	- \\
\hline \hline		
\end{tabular}
\caption{Parameter prior distributions. The starred values in brackets (*) for $\mathbb{E}[K(0)]$ denote the case where we consider an erroneous initial condition.}
\label{tab:priors}
\end{table}

\subsection{Case 1 (small change in stiffness)}\label{baseline}
\begin{figure}[H]
\centering
\includegraphics[width=\linewidth,keepaspectratio]{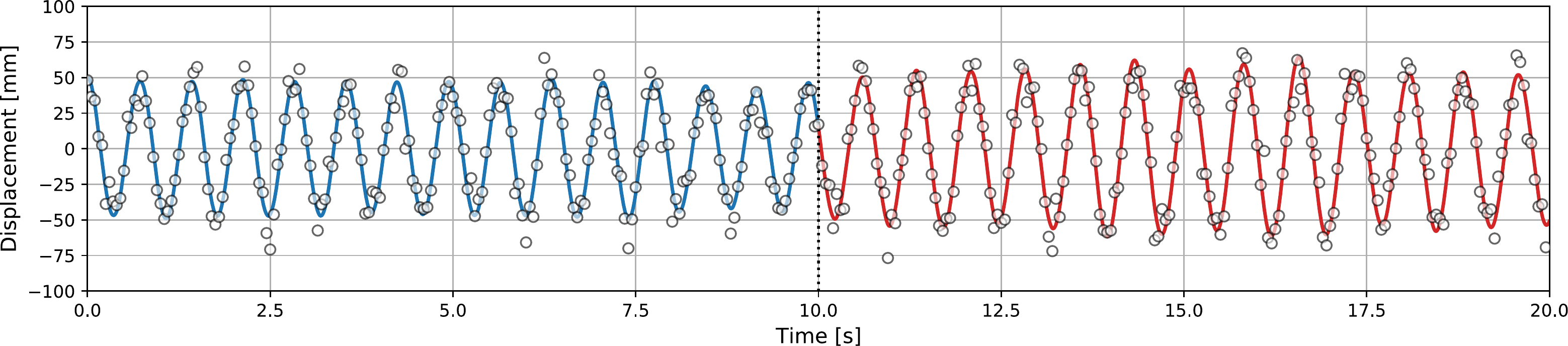}
\caption{Synthetic data for Case 1}
\label{fig:case1}
\end{figure}

The first case using the generated data will represent a baseline case. From there we will study the robustness of the proposed approach for different magnitudes of the sudden change in parameter value, and with respect to the noise strength and sparsity of the measurements. Figure \ref{fig:case1} shows a realization of the system (solid line), and the synthetic data points that are generated from it (white circles). The blue colour in the figure indicates that the system is in its undamaged state (having a stiffness of 80 N/mm), and the red colour shows the displacement in the system's damaged state (having a stiffness of 70 N/mm).

\subsubsection{Model 1}
\begin{figure}[H]
\centering
\includegraphics[width=\linewidth,keepaspectratio]{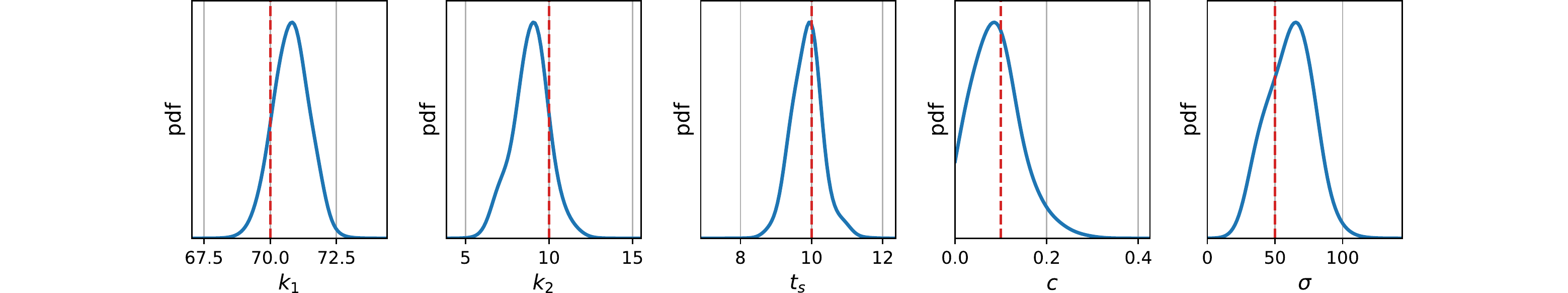}
\caption{Parameter posterior estimates of $\mathcal{M}_1$}\label{fig:c1m1pdf}
\end{figure}

Model $\mathcal{M}_1$ in the candidate set is identical to the data-generating model in Eq. (\ref{eq:mck_dynamics}) and Eq. (\ref{eq:gen12}). The marginal parameter posterior estimates are shown in Figure \ref{fig:c1m1pdf}. For comparison, the parameter values used to generate the synthetic data are identified by a dotted line.  Despite the change in dynamics being subtle visually, the parameter posterior estimates obtained by the TMCMC algorithm capture the true parameter values.

Although not shown here for brevity, the joint distribution of $k_1$ and $k_2$ shows that the parameters are highly correlated with the samples from the posterior being concentrated around the line $k_1 + k_2 = 80$.

\subsubsection{Model 2}

\begin{figure}[H]
\centering
\includegraphics[width=\linewidth,keepaspectratio]{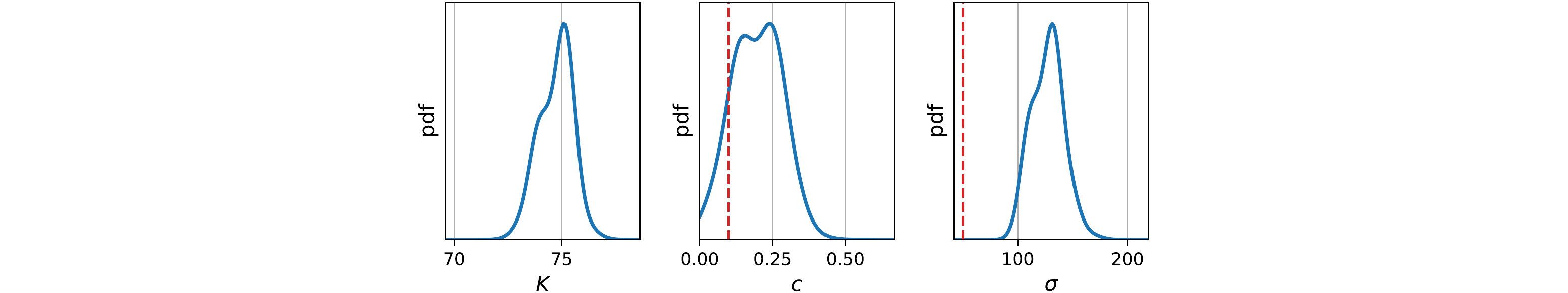}
\caption{Parameter posterior estimates of $\mathcal{M}_2$}\label{fig:c1m2pdf}
\end{figure}

Model $\mathcal{M}_2$ from the candidate set has the same form as the generating model in Eq. (\ref{eq:mck_dynamics}), but the model of the stiffness parameter is modelled as a single static parameter and does not account for any piece-wise change in stiffness in time. As seen in Figure \ref{fig:c1m2pdf}, the posterior distribution of the stiffness parameter is essentially bounded by the the stiffness values of 80 N/mm (before the snap) and 70 N/mm (after the snap), with near-zero probability outside these values. The reduced flexibility of this model forces the parameter estimate of the stiffness to be intermediate of the two true values, which is complemented by an overestimate of the stochastic forcing term, which here plays the role of capturing the model error strength. The increase in this parameter leads to an increase in the uncertainty in the state estimates during the forecast steps where no data are available.

\subsubsection{Model 3}
Model $\mathcal{M}_3$ once again models all parameters as static, however the stochastic forcing term $f(t)$ is now represented by a coloured noise process as seen in Eq. (\ref{eq:candidate3}) whereas in Eq. (\ref{eq:candidate2}) it is a white noise process. The coloured noise model is given by the Ornstein-Uhlenbeck process \cite{uhlenbeck1930theory}. The time correlation of the coloured noise (captured by the parameter $\tau$) introduces an additional parameter to be estimated, and also augments the state vector, to include the forcing term as it is now defined according to a first order It\^o differential equation. The use of a coloured noise model for the model error strength to capture model discrepancy is motivated by a previous study by Bisaillon et al. \cite{bisaillon2022combined}, which compared the performance of white noise and coloured noise processes for capturing model error for similar problems in nonlinear aeroelasticity. The state space model for this system with a time-correlated model error is given by \cite{gillespie1996exact}

\begin{subequations}
\begin{align}
\mathbf{x}_{1,k+1} &= \mathbf{x}_{1,k} + \Delta t \mathbf{x}_{2,k}, \\
\mathbf{x}_{2,k+1} &= (-\Delta t \ k /m) \mathbf{x}_{1,k}  + (1 - \Delta t \ c/m) \mathbf{x}_{2,k} + (\Delta t /m)  \mathbf{x}_{3,k}, \\
\mathbf{x}_{3,k+1} &= (1 - \Delta t/\tau)\mathbf{x}_{3,k} + \sqrt{\Delta t} \sigma \ \xi_{k}. \label{eq:model3c}
\end{align}
\end{subequations}

\begin{figure}[H]
\centering
\includegraphics[width=\linewidth,keepaspectratio]{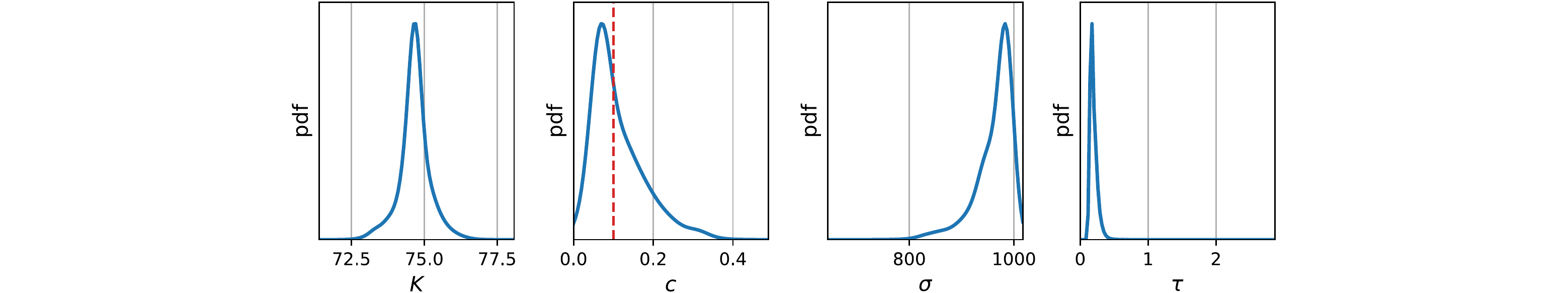}
\caption{Parameter posterior estimates of $\mathcal{M}_3$}\label{fig:c1m3pdf}
\end{figure}

\noindent
The marginal parameter posterior pdfs are shown in Figure \ref{fig:c1m3pdf}. The estimate of the stiffness parameter is similar to what was observed for model $\mathcal{M}_2$. The physical interpretation of the noise strength parameter $\sigma$ is different in this case, as the white noise process forces the system indirectly through the first-order differential equation for $f(t)$ as in Eq. (\ref{eq:model3c}), unlike the other cases where this white noise process forces the system directly.

\subsubsection{Model 4}
\begin{figure}[H]
\centering
\begin{subfigure}{0.475\linewidth}
\includegraphics[width=\linewidth,keepaspectratio]{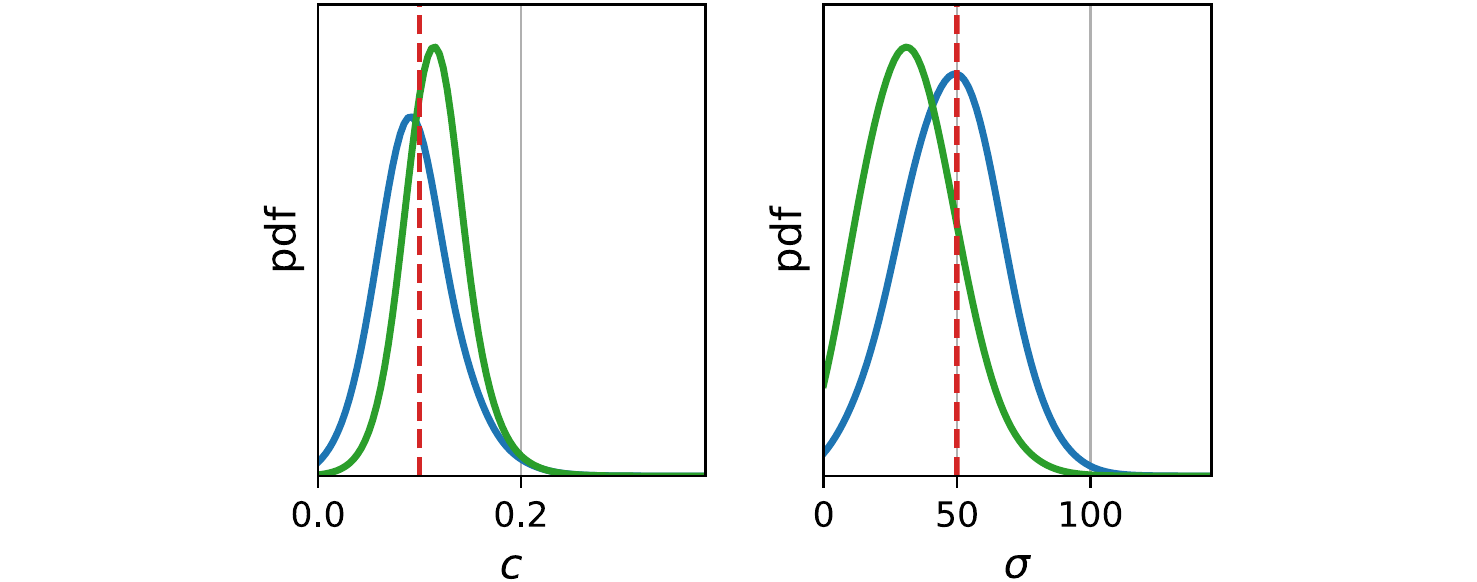}
\caption{Static parameter posterior estimates}\label{fig:c1m4apdf}
\end{subfigure}
\begin{subfigure}{0.475\linewidth}
\includegraphics[width=\linewidth,keepaspectratio]{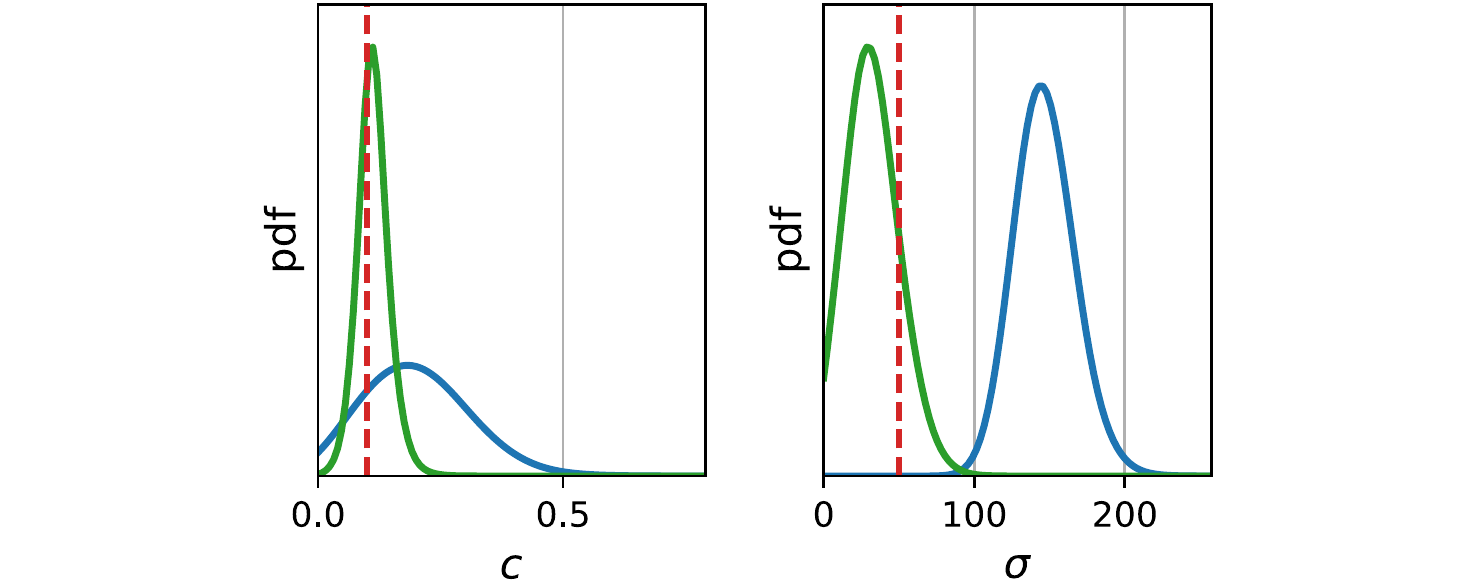}
\caption{Static parameter posterior estimates}\label{fig:c1m4bpdf}
\end{subfigure}
\begin{subfigure}{\linewidth}
\centering
\includegraphics[width=\linewidth,keepaspectratio]{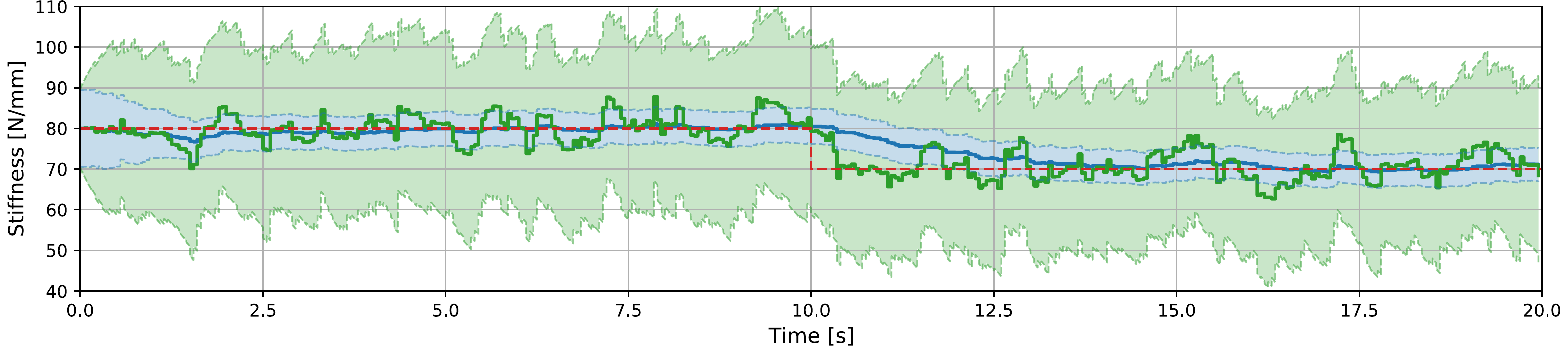}
\caption{EKF estimated mean (solid lines) $\pm$ 3 standard deviations (shaded area) for $K(t)$ at MAP estimates of static parameters}\label{fig:c1m4ase}
\end{subfigure}
\begin{subfigure}{\linewidth}
\centering
\includegraphics[width=\linewidth,keepaspectratio]{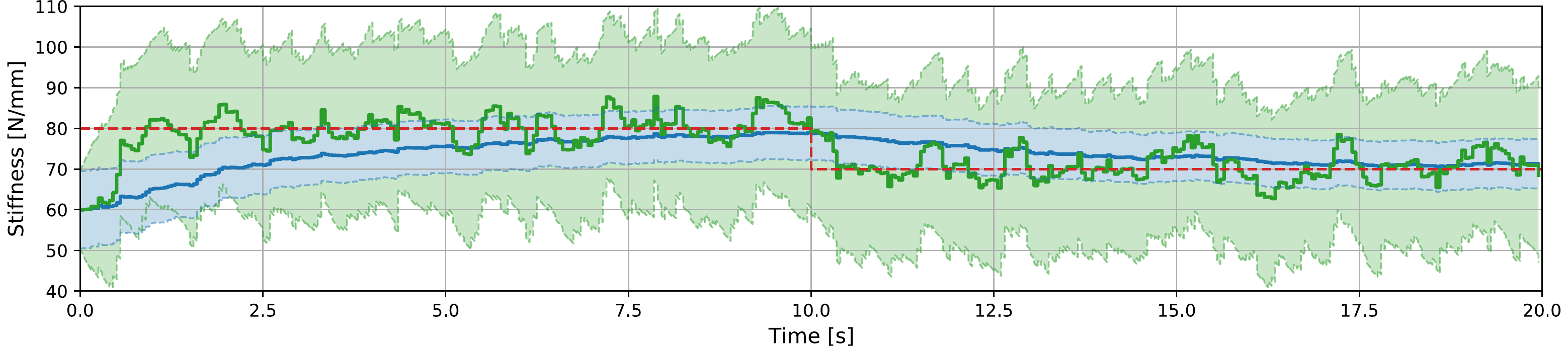}
\caption{EKF estimated mean (solid lines) $\pm$ 3 standard deviations (shaded area) for $K(t)$ at MAP estimates of static parameters}\label{fig:c1m4bse}
\end{subfigure}
\caption{Parameter estimates of $\mathcal{M}_4^a$ (blue) and $\mathcal{M}_4^b$ (green), and true parameter values (red), with initial mean $\mathbb{E}[K(0)] = 80$ (panels (\ref{fig:c1m4apdf}) and (\ref{fig:c1m4ase})), and with initial mean $\mathbb{E}[K(0)] = 60$ (panels (\ref{fig:c1m4bpdf}) and (\ref{fig:c1m4bse}))}\label{fig:c1m4all}
\end{figure}

The first three candidate models only considered system parameters as static. We now model the system stiffness as a time-varying parameter, with the remaining parameters treated as time-invariant. This is accomplished by augmenting the state-space model, introducing a third state which represents the stiffness, where its velocity is modelled according to a zero-mean Gaussian random variable. In this model, the state space model is no longer linear

\begin{subequations}
\begin{align}
\mathbf{x}_{1,k+1} &= \mathbf{x}_{1,k} + \Delta t \mathbf{x}_{2,k}, \\
\mathbf{x}_{2,k+1} &= (-\Delta t /m) \mathbf{x}_{1,k}\mathbf{x}_{3,k}  + (1 - \Delta t c/m) \mathbf{x}_{2,k} + \sqrt{\Delta t} \ \sigma \xi_{1,k}/m, \\
\mathbf{x}_{3,k+1} &= \mathbf{x}_{3,k} + \sqrt{\Delta t} \gamma \ \xi_{2,k}. \label{eq:model3c}
\end{align}
\end{subequations}

where $\xi_{1,k}$ and $\xi_{2,k}$ are independent unit standard Gaussian random variables (i.e., having zero mean and unit variance) and $\mathbf{x}_k = \{\mathbf{x}_{1,k}, \mathbf{x}_{2,k}, \mathbf{x}_{3,k}\}^{\text{T}} = \{u_k, \dot{u}_k, K_k\}^{\text{T}}$. In a similar fashion to the noise strength parameter for the random forcing, $\sigma$, the parameter $\gamma$ introduced here, controls the strength of the artificial noise that drives the dynamics of the stiffness parameter. Model $\mathcal{M}_4$ considers this parameter to be known a priori, as if it had been manually tuned, and thus does not attempt to estimate it. In the first scenario, $\gamma$ assumes a value of $\gamma_a = 1$ (labelled as $\mathcal{M}_4^a$), and in the second scenario, $\gamma_b = 10$  (labelled as $\mathcal{M}_4^b$). \textcolor{oct15}{The results for $\gamma_a = 1$ are shown in blue, and for $\gamma_b = 10$ are shown in green in Figure \ref{fig:c1m4all}. Figures \ref{fig:c1m4apdf} and \ref{fig:c1m4ase}, show the marginal posterior distributions of the static parameters and transitional pdfs of the time-varying stiffness parameter where the initial mean of the time-varying stiffness is the true value of 80 N/mm. Subsequently, we see how the results differ in the case where the initial mean is incorrect in Figures \ref{fig:c1m4bpdf} and \ref{fig:c1m4bse}. 
For the case where the initial stiffness is correct, the state estimation results in Figure \ref{fig:c1m4ase} show that the change in stiffness is tracked reasonably well by both models $\mathcal{M}_4^a$ and $\mathcal{M}_4^b$. Model $\mathcal{M}_4^a$ has much lower uncertainty in the state estimates, however it also adjusts to the change in stiffness more gradually, whereas $\mathcal{M}_4^b$ adapts to the change much more quickly, but the state estimates in time are also much more irregular. The effect of this on the marginal posterior estimates of the damping and stochastic forcing terms can be seen in Figure \ref{fig:c1m4apdf}. The the modes of the pdfs match the true parameter values for model $\mathcal{M}_4^a$, whereas the strength of the stochastic forcing is slightly underestimated in $\mathcal{M}_4^b$ as the randomness in the forcing is being accounted for in the highly oscillatory nature of the estimates of the stiffness. 
When the initial mean of the stiffness parameter is not known correctly, it can be seen in Figure \ref{fig:c1m4ase} that model $\mathcal{M}_4^a$ takes a long time to correct for the erroneous initial condition, leading to poor esitmates of the stiffness over the majority of the period of observation. Due to the large noise strength in $\mathcal{M}_4^b$, the results shown are relatively insensitive to the initial condition as it quickly adjusts to the correct value. The posterior pdfs in Figure \ref{fig:c1m4bpdf} are  relatively insensitive to this change for model $\mathcal{M}_4^b$, whereas for model $\mathcal{M}_4^b$ it can be seen that the posterior pdf for the damping parameter is subject to much more uncertainty. The stochastic forcing strength is over-estimated as the model error is relied on to account for the discrepancy between the observed response and the predicted response based on the incorrect estimates of the stiffness parameter.}

\subsubsection{Model 5}
\begin{figure}[H]
\centering
\begin{subfigure}{0.475\linewidth}
\includegraphics[width=\linewidth,keepaspectratio]{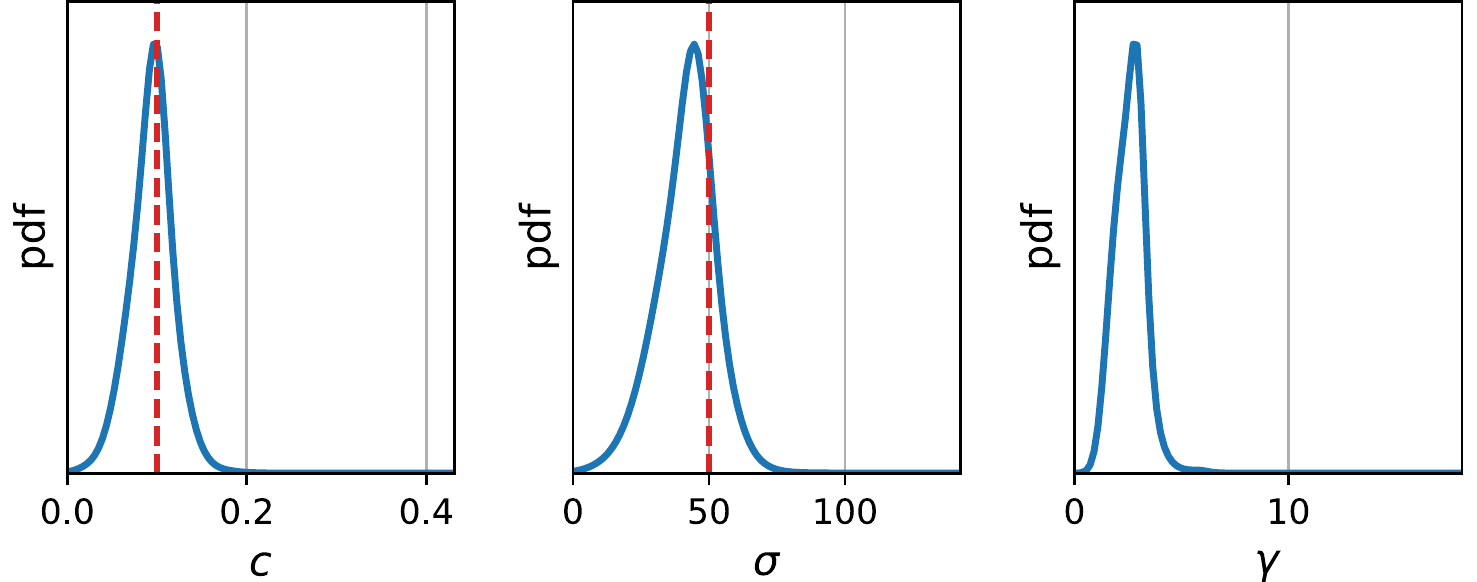}
\caption{Static parameter posterior estimates}\label{fig:c1m5apdf}
\end{subfigure}
\begin{subfigure}{0.475\linewidth}
\includegraphics[width=\linewidth,keepaspectratio]{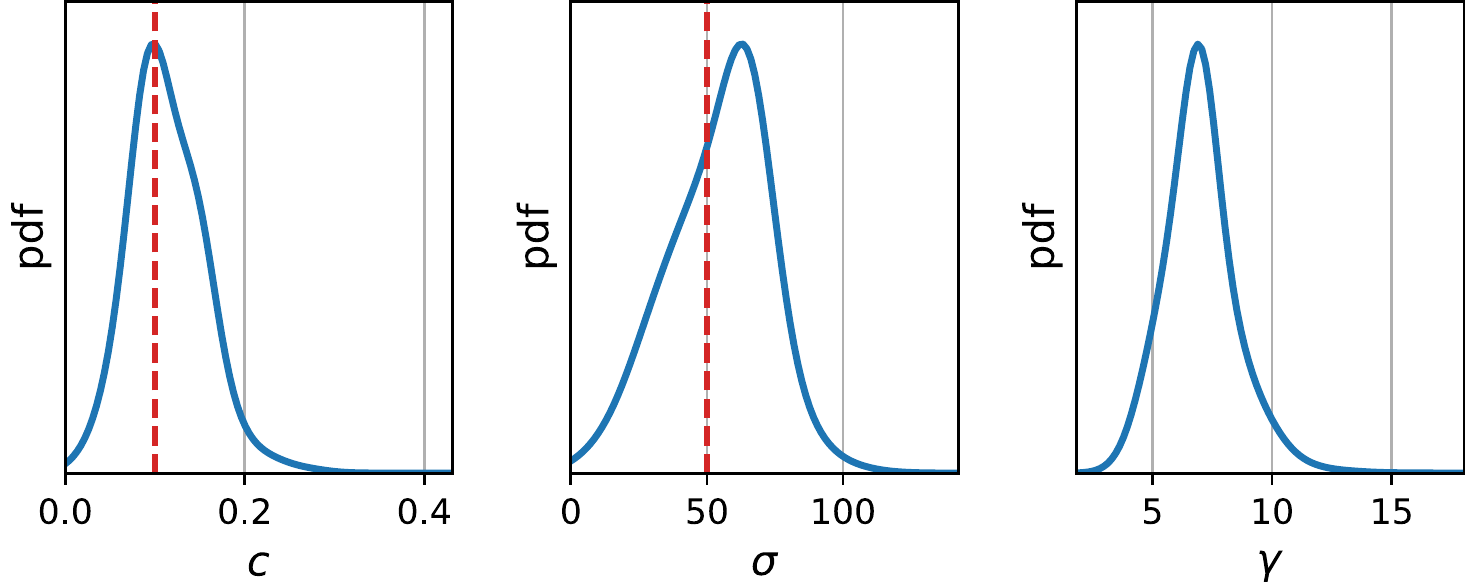}
\caption{Static parameter posterior estimates}\label{fig:c1m5bpdf}
\end{subfigure}
\begin{subfigure}{\linewidth}
\centering
\includegraphics[width=\linewidth,keepaspectratio]{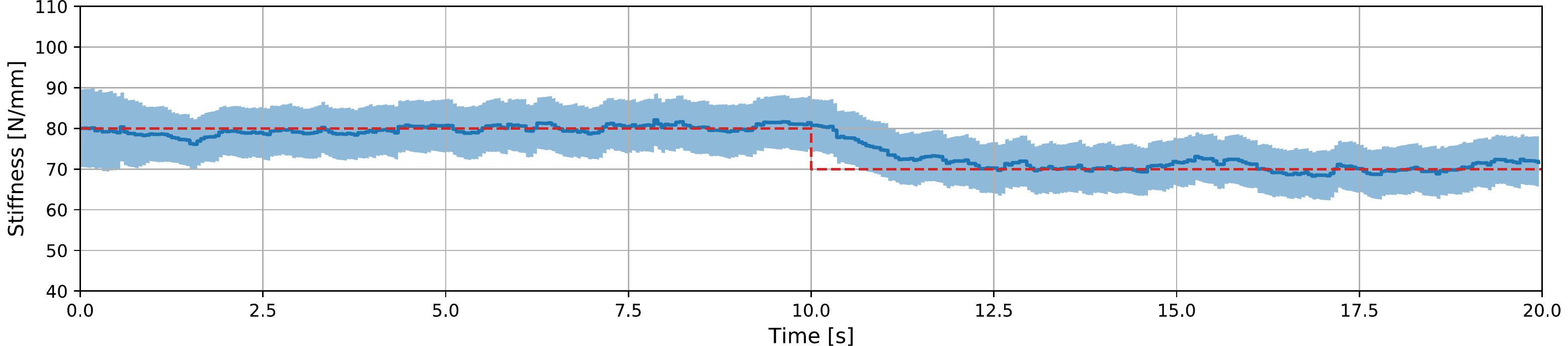}
\caption{EKF estimated mean (solid lines) $\pm$ 3 standard deviations (shaded area) for $K(t)$ at MAP estimates of static parameters}\label{fig:c1m5ase}
\end{subfigure}
\begin{subfigure}{\linewidth}
\centering
\includegraphics[width=\linewidth,keepaspectratio]{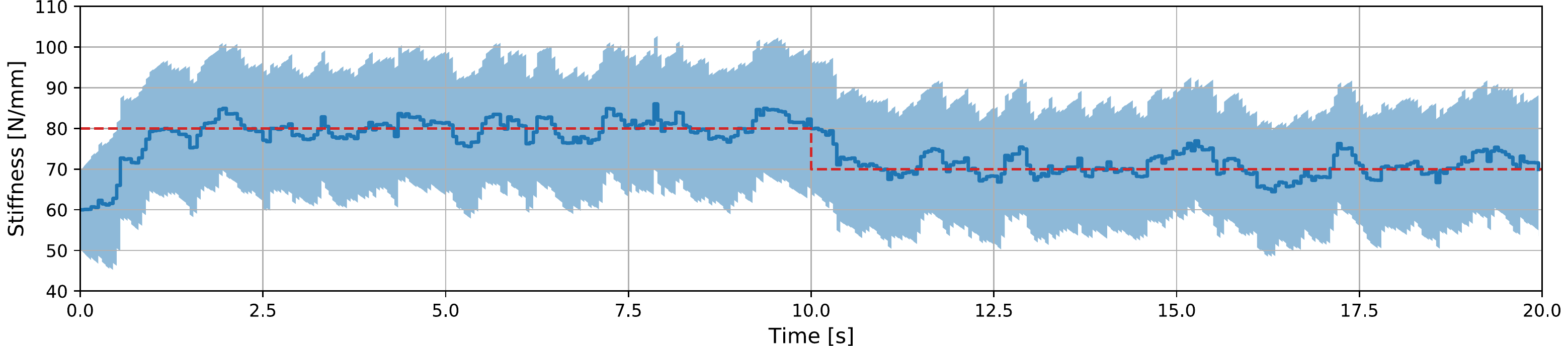}
\caption{EKF estimated mean (solid lines) $\pm$ 3 standard deviations (shaded area) for $K(t)$ at MAP estimates of static parameters}\label{fig:c1m5bse}
\end{subfigure}
\caption{Parameter estimates of $\mathcal{M}_5$ (blue) and true parameter values (red), with initial mean $\mathbb{E}[K(0)] = 80$ (panels (\ref{fig:c1m5apdf}) and (\ref{fig:c1m5ase})), and with initial mean $\mathbb{E}[K(0)] = 60$ (panels (\ref{fig:c1m5bpdf}) and (\ref{fig:c1m5bse}))}\label{fig:c1m5all}
\end{figure}

Model $\mathcal{M}_5$ has the same form as model $\mathcal{M}_4$, however it introduces additional flexibility as the value of the artificial noise strength, $\gamma$, in Eq. (\ref{eq:model3c}) is introduced as a static parameter to be estimated through MCMC. The estimation of this parameter alongside the system parameters automates the tuning of the noise strength parameter driving the stiffness parameter. {The marginal posterior distributions of the static parameters and transitional pdfs of the time-varying stiffness parameter with the correct initial mean stiffness are shown in Figures \ref{fig:c1m5apdf} and \ref{fig:c1m5ase}. The same results where the initial mean stiffness is incorrect are shown in Figures \ref{fig:c1m5bpdf} and \ref{fig:c1m5bse}. Note that due to the small change in stiffness at 10s, when the initial conditions are correct, the optimal noise strength perturbing the stiffness parameter is relatively small, resulting in estimates with low uncertainty, and gradual changes over time. When the initial conditions are incorrect, a higher noise strength is needed to perturb the stiffness parameter, which allows for the stiffness estimates to automatically correct themselves more quickly, but result in more oscillatory parameter estimates within higher uncertainty.}

\subsubsection{Model 6}
\begin{figure}[H]
\centering\centering
\begin{subfigure}{\linewidth}
\includegraphics[width=\linewidth,keepaspectratio]{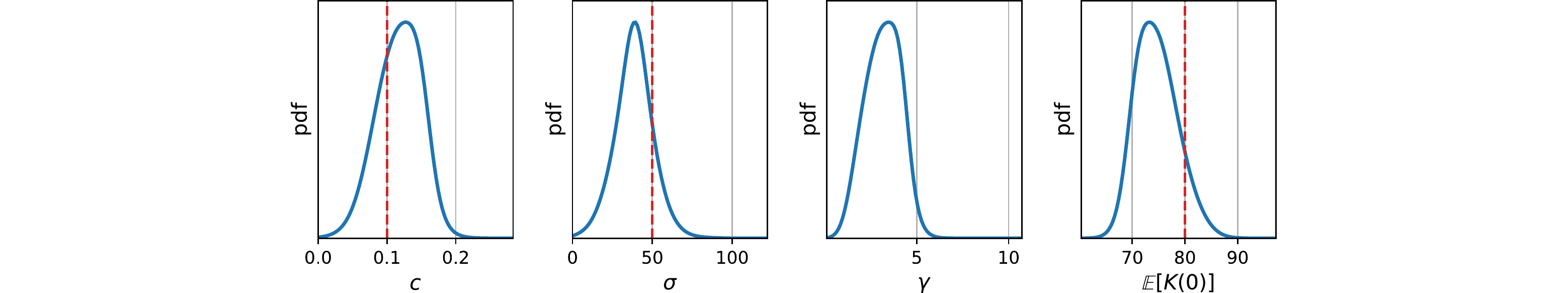}
\caption{Static parameter posterior estimates}\label{fig:c1m6pdf}
\end{subfigure}
\begin{subfigure}{\linewidth}
\centering
\includegraphics[width=\linewidth,keepaspectratio]{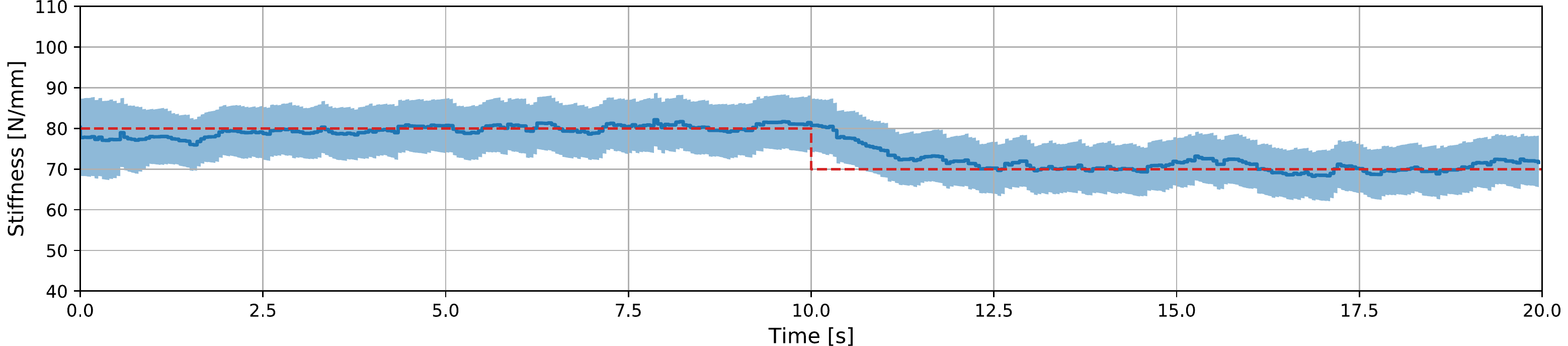}
\caption{EKF estimated mean (solid lines) $\pm$ 3 standard deviations (shaded area) for $K(t)$ at MAP estimates of static parameters}\label{fig:c1m6se}
\end{subfigure}
\caption{Parameter estimates of $\mathcal{M}_6$ (blue) and true parameter values (red)}\label{fig:c1m6all}
\end{figure}

Model $\mathcal{M}_6$ has the same form as models $\mathcal{M}_4$ and $\mathcal{M}_5$, but introduces further flexibility as both $\gamma$ and the mean initial stiffness $\mathbb{E}[K(0)]$ are treated as time-invariant parameters to be estimated. \textcolor{oct15}{The marginal posterior distributions of the static parameters and transitional pdfs of the time-varying stiffness parameter are shown in Figure \ref{fig:c1m6all}. The ability to estimate the initial mean of the stiffness parameter alleviates the need to correct erroneous initial conditions, and thus the noise strength $\gamma$ is dictated by the amplitude of the degradation in stiffness, rather than by the initial conditions.}

\subsubsection{Model selection results and discussion}\label{c1_modelsel}
Table \ref{table:case1evidence} summarizes the results of the model comparison as outlined in Section \ref{section:modelsel}. Though the model selection component is not the ultimate goal of the paper, we use it as a basis for comparing the relative performance of the various proposed models. Particularly to assess the performance of the models consisting of a combination of static parameters (estimated using MCMC) and time-varying parameters (estimated concurrently using nonlinear filters). 

\begin{table}[!h]
{\footnotesize
\centering
\begin{tabular}{lrrrrrrrrrr}
\hline
 \Big. Model & $\mathcal{M}_1$ & $\mathcal{M}_2$ & $\mathcal{M}_3$ & \multicolumn{2}{c}{${\mathcal{M}_4}^a$} &  \multicolumn{2}{c}{${\mathcal{M}_4}^b$} & \multicolumn{2}{c}{$\mathcal{M}_5$} & $\mathcal{M}_6$ \\
\hline
\Big. log Evidence (-1600) 	&75.45	&54.21	&43.05	&77.28	&(60.79)	&68.19	&(64.81)&	74.50	&(63.66)&	72.27  \\
 \Big. Average data-fit (-1600) 	& 98.41&	65.51&	58.30&	83.59&	(66.81)&	74.82&	(71.36)&	87.25&	(75.28)&	86.94\\
 \Big. Expected information gain &22.96	&11.30	&15.24	&6.32	&(6.03)&	6.63	&(6.55)	&12.74	&(11.62)&	14.67
\\
\hline
 \Big. Model probability (\%) &	13.06&	0.00&	0.00&	81.31&	(0.00)&	0.01&	(0.00)&	5.08&	(0.00)&	0.54 \\	
\hline
\hline									
\end{tabular}
\caption{Model selection results for Case 1. Bracketed values correspond to poor estimates of the initial stiffness.}\label{table:case1evidence}
}
\end{table}

Here, model $\mathcal{M}_4^a$ has the highest model probability despite not having the best average data-fit. Models $\mathcal{M}_1$, $\mathcal{M}_5$ and $\mathcal{M}_6$, each have superior data-fitting capabilities (listed in descending order), however, are penalized by their respective expected information gain, which is used as a metric for model complexity.  Model $\mathcal{M}_1$ has an average data-fit that is significantly larger than the models that follow, which is perhaps intuitive, as it has the same structure as the generating model. Comparatively, models $\mathcal{M}_2$ and $\mathcal{M}_3$ which do not consider any variation in the system stiffness, and instead rely on the model error/stochastic forcing (white and coloured noise, respectively) to account for the unmodelled physics, exhibit poor data-fit.
Note that the expected information gain will depend on the number of static parameters and the support of the uniform prior distributions listed in Table \ref{tab:priors}. The support of the parameters are made arbitrarily large to be non-informative. However, in the case where the parameter prior distributions are informative, the KL-divergence of the posterior distributions from the prior distributions may be reduced, and thus the penalty assessed to these more complex models may be lessened.

\subsection{Case 2 (large change in stiffness)}\label{sudden}
\begin{figure}[H]
\centering
\includegraphics[width=\linewidth,keepaspectratio]{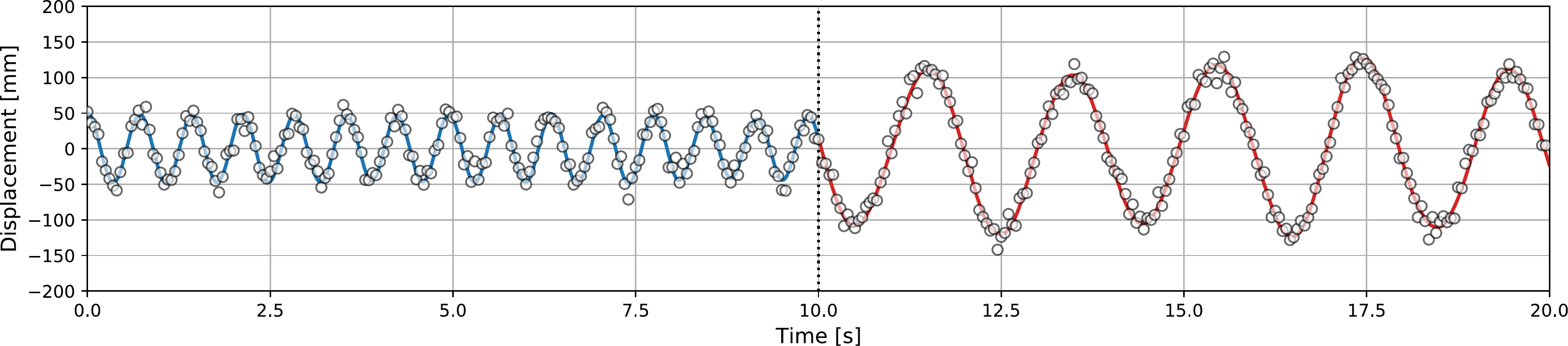}
\caption{Synthetic data for Case 2}
\label{fig:case2}
\end{figure}

In Case 1 above, the spring having the lower stiffness value snapped at time $t=t_s$, inducing a step-wise 12.5\% decrease in the overall stiffness (from 80 N/mm to 70N/mm). This small change in stiffness allowed us to assess the various modelling techniques abilities to detect this change through noisy observations of the system response that was visually imperceptible. In Case 2, we test the algorithm against the scenario where the stiffer of the two springs snaps at time $t = t_s$, resulting in a much more significant reduction in the overall stiffness (from 80 N/mm to 10 N/mm). In this case, the change is visually noticeable, and instead of testing the algorithm's ability to detect subtle changes, we assess its ability to adapt to significant and abrupt changes. Figure \ref{fig:case2} shows a realization of the system, and the synthetic data points that are generated from it. The blue colour for the true signal indicates that the system is in its undamaged state prior to the snap (having a stiffness of 80 N/mm), whereas beyond the snap the colour red shows the system's response in its damaged state (having a stiffness of 10 N/mm).

\begin{figure}[H]
\centering
\includegraphics[width=\linewidth,keepaspectratio]{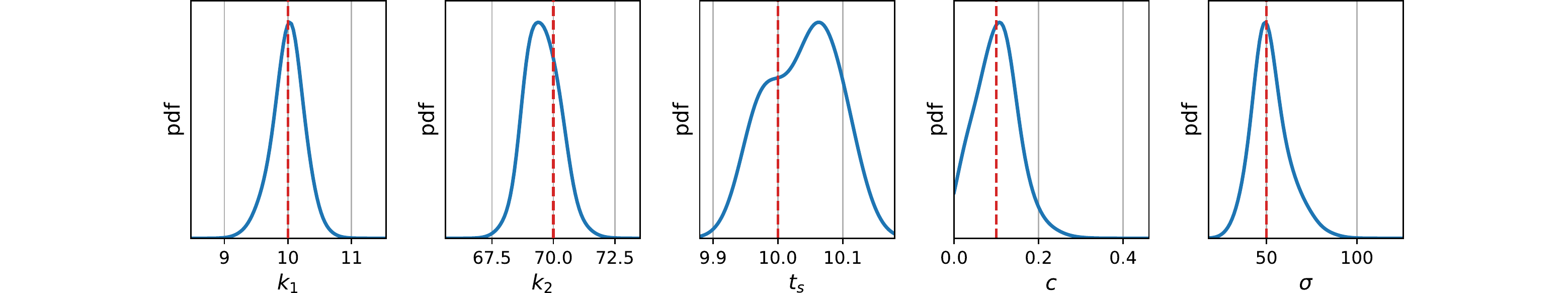}
\caption{Parameter posterior estimates of $\mathcal{M}_1$}\label{fig:c2m1pdf}
\end{figure}

\begin{figure}[H]
\centering
\includegraphics[width=\linewidth,keepaspectratio]{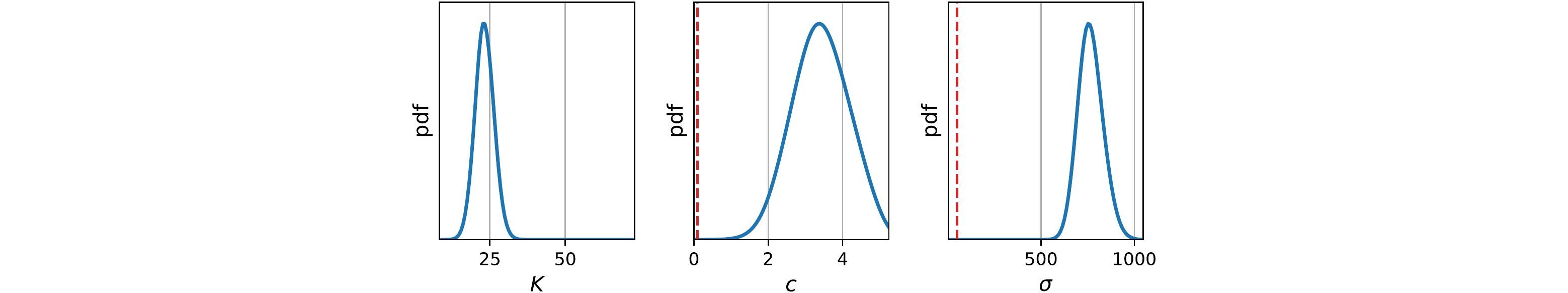}
\caption{Parameter posterior estimates of $\mathcal{M}_2$}\label{fig:c2m2pdf}
\end{figure}

\begin{figure}[H]
\centering
\includegraphics[width=\linewidth,keepaspectratio]{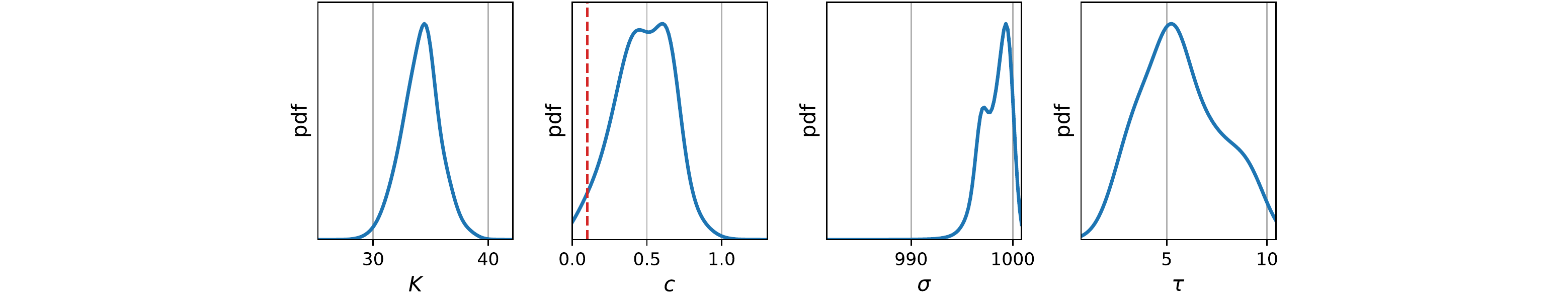}
\caption{Parameter posterior estimates of $\mathcal{M}_3$}\label{fig:c2m3pdf}
\end{figure}

\textcolor{oct15}{The parameter posterior pdfs for models $\mathcal{M}_1$, $\mathcal{M}_2$, and $\mathcal{M}_3$, are reported in Figures \ref{fig:c2m1pdf}, \ref{fig:c2m2pdf}, and \ref{fig:c2m3pdf}, respectively. Intuitively, as model $\mathcal{M}_1$ has the same form as the data-generating model, it performs well, and the true parameter values are well-captured by the posterior pdfs. With the significant change in stiffness, the form of models $\mathcal{M}_2$ and $\mathcal{M}_3$, make them inadequate for capturing the dynamics, as their posterior pdfs do not capture the true parameter values within their high probability density region, and the model instead relies principally on the model error to forecast between measurements.}

\begin{figure}[H]
\centering
\begin{subfigure}{0.475\linewidth}
\includegraphics[width=\linewidth,keepaspectratio]{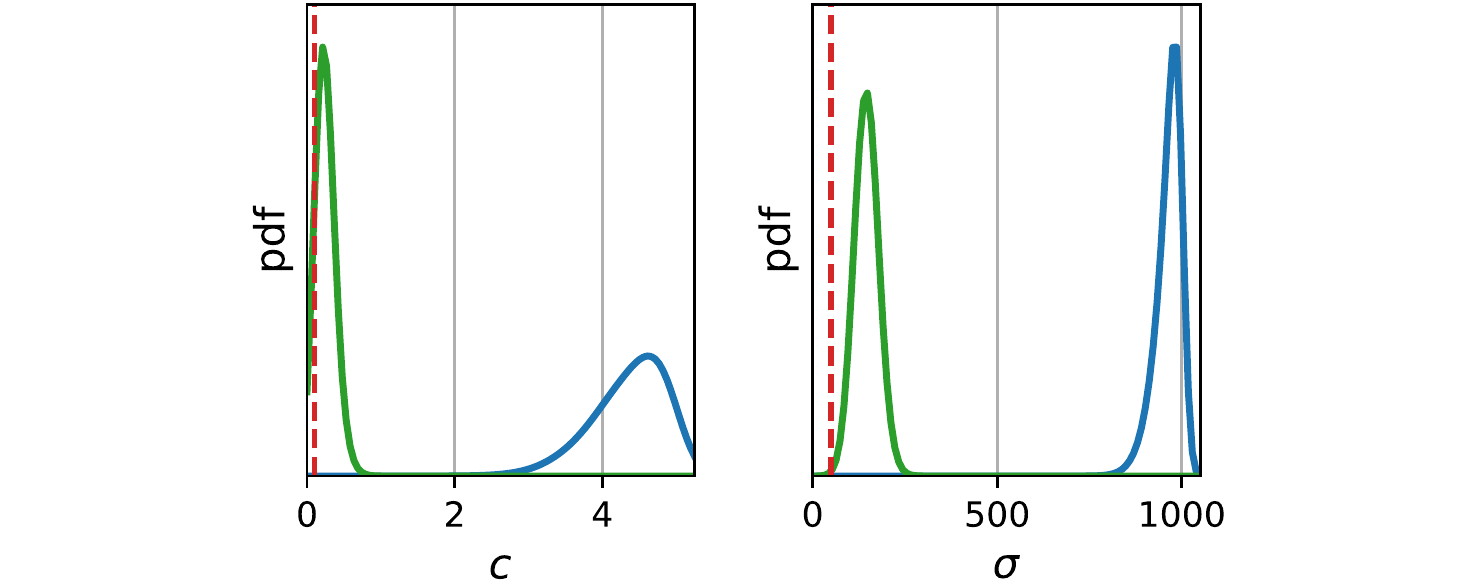}
\caption{Static parameter posterior estimates}\label{fig:c2m4apdf}
\end{subfigure}
\begin{subfigure}{0.475\linewidth}
\includegraphics[width=\linewidth,keepaspectratio]{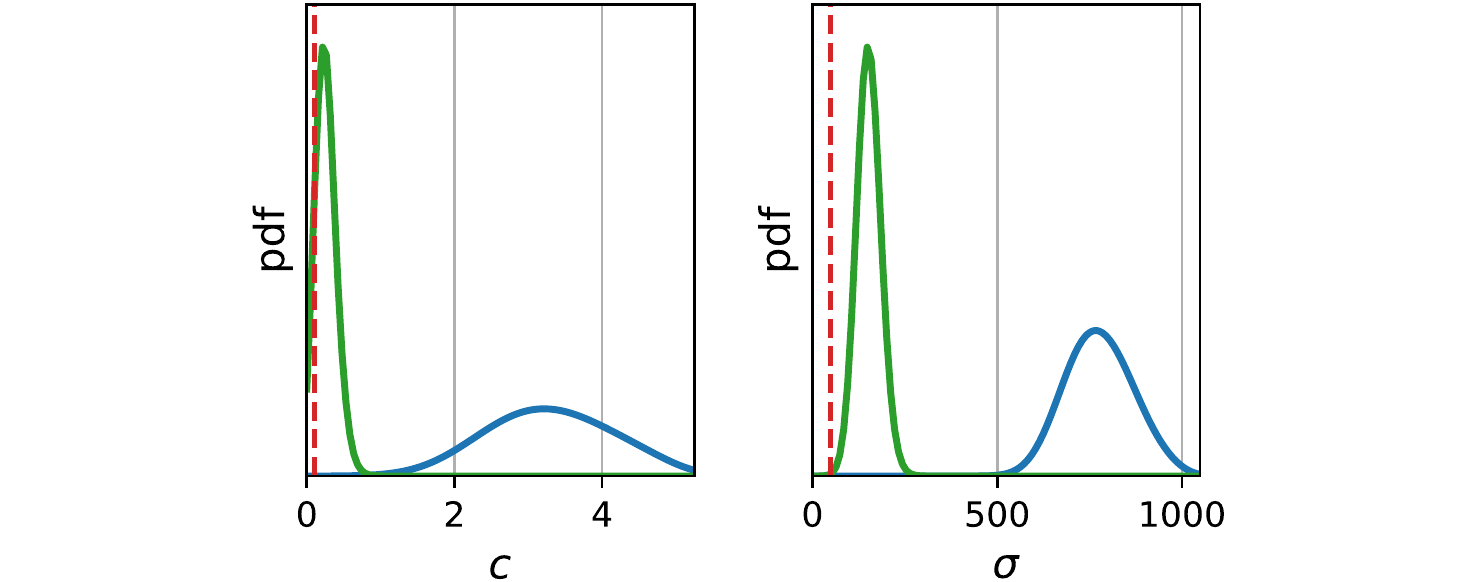}
\caption{Static parameter posterior estimates}\label{fig:c2m4bpdf}
\end{subfigure}
\begin{subfigure}{\linewidth}
\centering
\includegraphics[width=\linewidth,keepaspectratio]{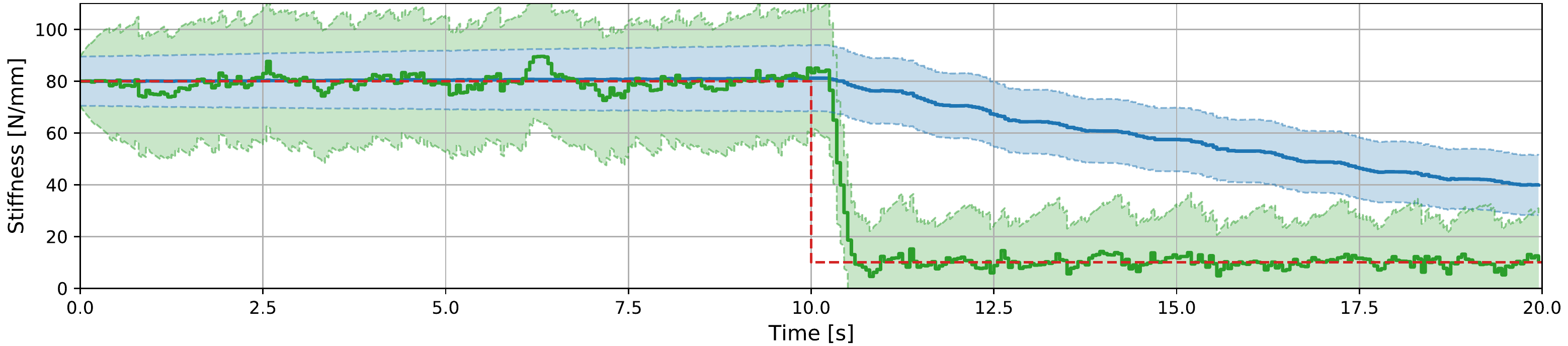}
\caption{EKF estimated mean (solid lines) $\pm$ 3 standard deviations (shaded area) for $K(t)$ at MAP estimates of static parameters}\label{fig:c2m4ase}
\end{subfigure}
\begin{subfigure}{\linewidth}
\centering
\includegraphics[width=\linewidth,keepaspectratio]{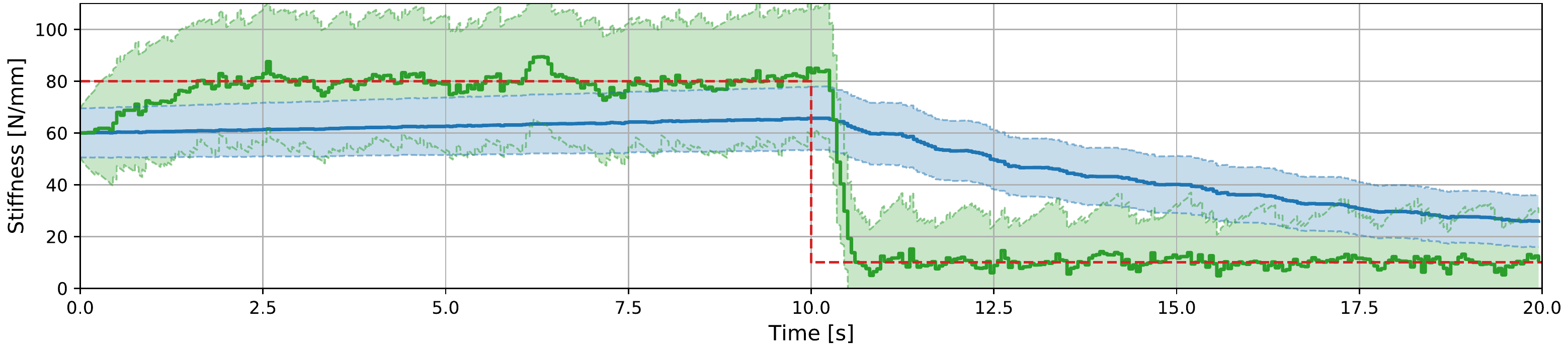}
\caption{EKF estimated mean (solid lines) $\pm$ 3 standard deviations (shaded area) for $K(t)$ at MAP estimates of static parameters}\label{fig:c2m4bse}
\end{subfigure}
\caption{Parameter estimates of $\mathcal{M}_4^a$ (blue) and $\mathcal{M}_4^b$ (green), and true parameter values (red), with initial mean $\mathbb{E}[K(0)] = 80$ (panels (\ref{fig:c2m4apdf}) and (\ref{fig:c2m4ase})), and with initial mean $\mathbb{E}[K(0)] = 60$ (panels (\ref{fig:c2m4bpdf}) and (\ref{fig:c2m4bse}))}\label{fig:c2m4}
\end{figure}

\begin{figure}[H]
\centering
\begin{subfigure}{0.475\linewidth}
\includegraphics[width=\linewidth,keepaspectratio]{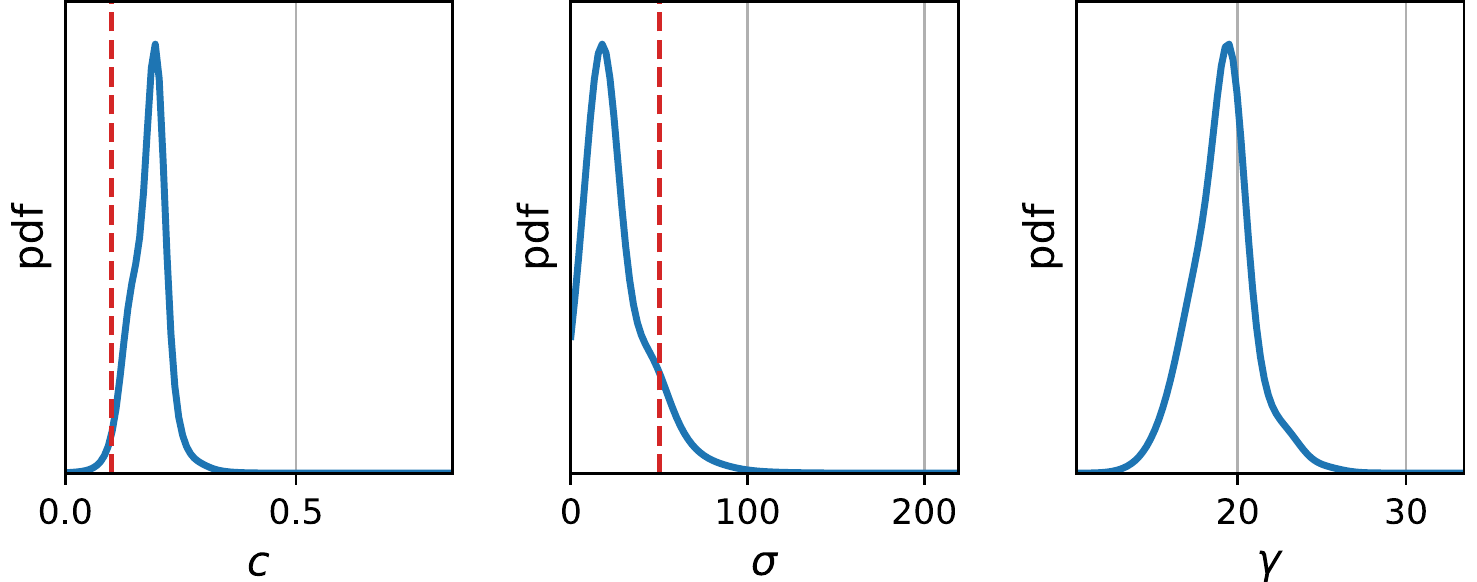}
\caption{Static parameter posterior estimates}\label{fig:c2m5apdf}
\end{subfigure}
\begin{subfigure}{0.475\linewidth}
\includegraphics[width=\linewidth,keepaspectratio]{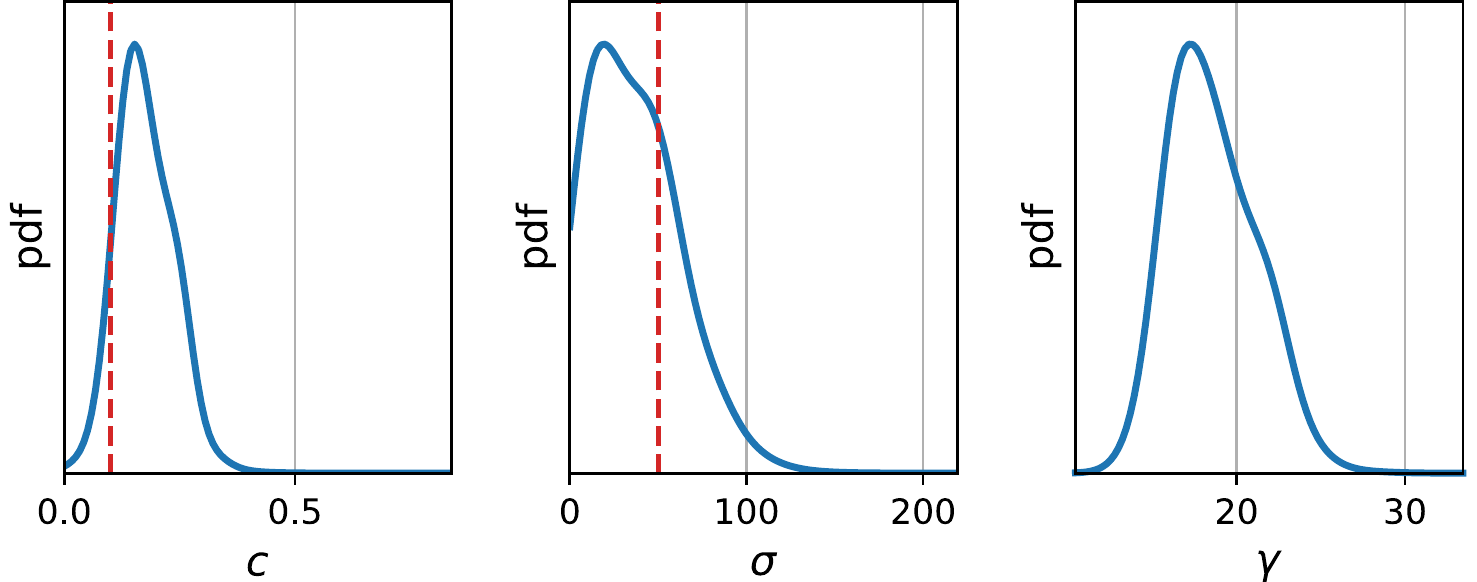}
\caption{Static parameter posterior estimates}\label{fig:c2m5bpdf}
\end{subfigure}
\begin{subfigure}{\linewidth}
\centering
\includegraphics[width=\linewidth,keepaspectratio]{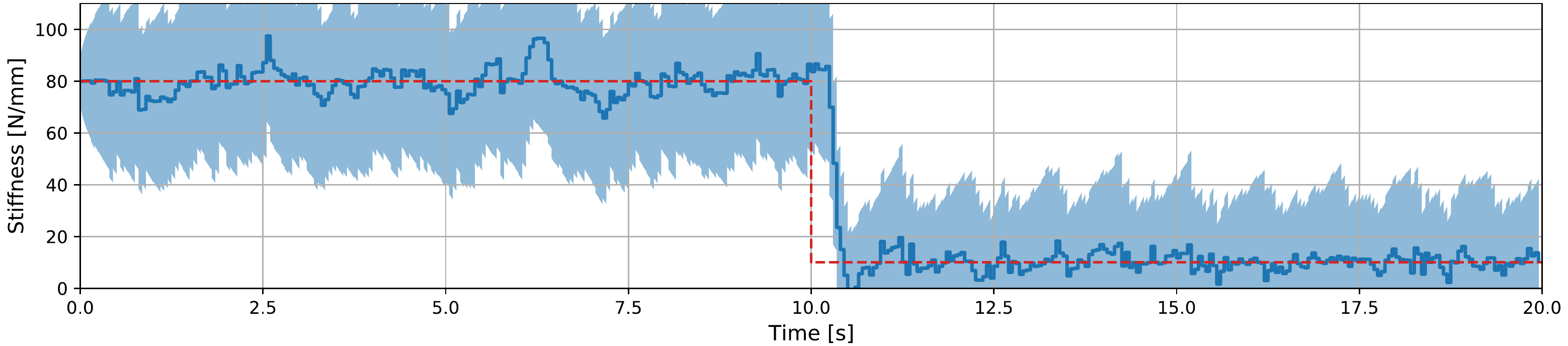}
\caption{EKF estimated mean (solid lines) $\pm$ 3 standard deviations (shaded area) for $K(t)$ at MAP estimates of static parameters}\label{fig:c2m5ase}
\end{subfigure}
\begin{subfigure}{\linewidth}
\centering
\includegraphics[width=\linewidth,keepaspectratio]{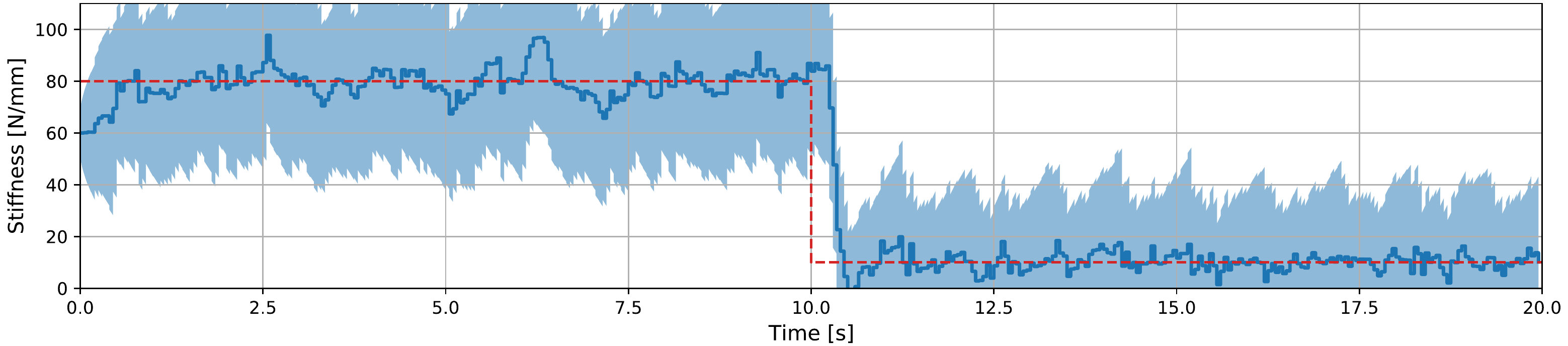}
\caption{EKF estimated mean (solid lines) $\pm$ 3 standard deviations (shaded area) for $K(t)$ at MAP estimates of static parameters}\label{fig:c2m5bse}
\end{subfigure}
\caption{Parameter estimates of $\mathcal{M}_5$ (blue) and true parameter values (red), with initial mean $\mathbb{E}[K(0)] = 80$ (panels (\ref{fig:c2m5apdf}) and (\ref{fig:c2m5ase})), and with initial mean $\mathbb{E}[K(0)] = 60$ (panels (\ref{fig:c2m5bpdf}) and (\ref{fig:c2m5bse}))}\label{fig:c2m5}
\end{figure}

\begin{figure}[H]
\centering
\begin{subfigure}{\linewidth}
\includegraphics[width=\linewidth,keepaspectratio]{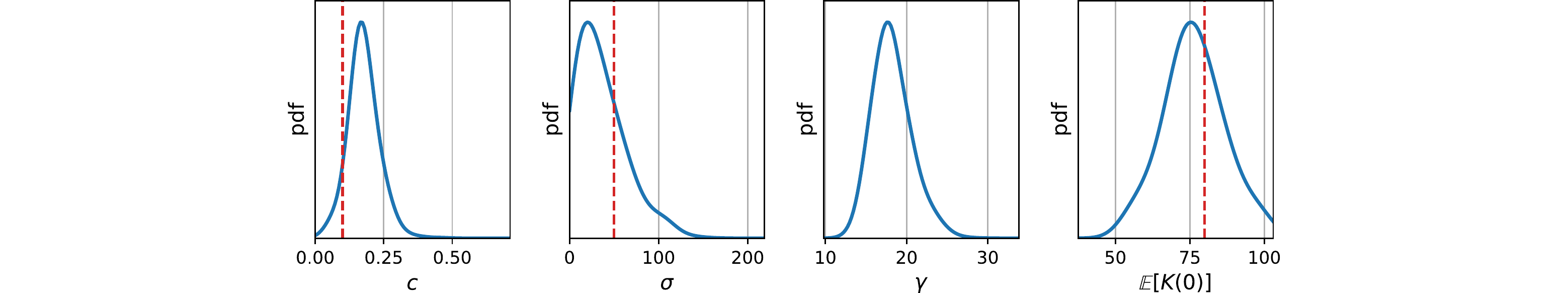}
\caption{Static parameter posterior estimates of $\mathcal{M}_6$}\label{fig:c2m6pdf}
\end{subfigure}
\begin{subfigure}{\linewidth}
\centering
\includegraphics[width=\linewidth,keepaspectratio]{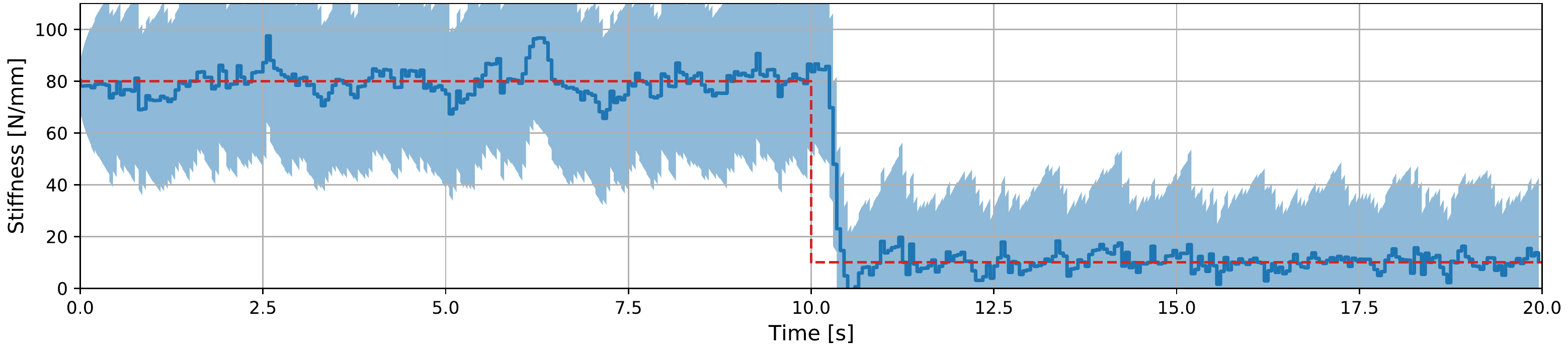}
\caption{EKF estimated mean (solid lines) $\pm$ 3 standard deviations (shaded area) for $K(t)$ at MAP estimates of static parameters}\label{fig:c2m6se}
\end{subfigure}
\caption{Parameter estimates of $\mathcal{M}_6$ (blue) and true parameter values (red)}\label{fig:c2m6}
\end{figure}

\textcolor{oct15}{We now turn our attention to the models which estimate a combination of time-varying and time-invariant parameters: $\mathcal{M}_4^a$, $\mathcal{M}_4^b$, $\mathcal{M}_5$, and $\mathcal{M}_6$. For models $\mathcal{M}_4^a$ and $\mathcal{M}_4^b$,  Figure \ref{fig:c2m4} illustrates the results when the artificial noise strength parameters had been tuned manually to $\gamma_a = 1$ and $\gamma_b =10$ respectively. In this case, $\gamma=1$ is clearly too small, as the parameter estimates are unable to track the significant and sudden change effectively. However, in this case, while the estimates look reasonable for $\gamma_b = 10$, this is still too small, as the stochastic forcing/model error strength parameter $\sigma$ is still over-estimated. The need to manually select a value for the artificial noise strength is alleviated by automatically estimating the parameter $\gamma$ by MCMC in model $\mathcal{M}_5$ shown in Figure \ref{fig:c2m5}, and by model $\mathcal{M}_6$ shown in Figure \ref{fig:c2m6}.}

\subsubsection{Model selection results and discussion}\label{c2_modelsel}
Table \ref{table:case2evidence} summarizes the results of the model comparisson for Case 2. As summarized in the first row of model proabilities, model $\mathcal{M}_1$, is shown to have 100\% probability, owing to its superior average data-fit resulting from having the same form as the generating model, and recovering all the true parameter values with relatively low uncertainty. 

In order to compare the remaining models, in the following row (denoted by $*$ in Table \ref{table:case2evidence}), we re-calculate the model probabilities after removing $\mathcal{M}_1$ from the candidate set. Models $\mathcal{M}_2$, and $\mathcal{M}_3$, which do not account for any change in stiffness, and instead rely on the model error to account for unmodelled physics perform poorly, as do both implementations of model $\mathcal{M}_4$. Recall from Table \ref{table:case1evidence}, that model $\mathcal{M}_4^a$ has the highest model probability in Case 1, due to its adequate data-fitting capability and its simplicity (resulting in a low penalty due to the expected information gain). In Case 2, however, it can be seen in Figure \ref{fig:c2m4} that the artificial noise strength driving the stiffness parameter, $\gamma_a = 1$ and $\gamma_b =10$ are too low, and delay the algorithm's ability to adjusting to the change in stiffness here. This has adverse effects on the data-fit of the model. Compare these fixed values of $\gamma_a$ and $\gamma_b$  to the parameter posterior estimates for $\gamma$ in models $\mathcal{M}_5$ (Figure \ref{fig:c2m5}), and $\mathcal{M}_6$ (Figure \ref{fig:c2m6}), where all of the mass is concentrated at values of $\gamma$ greater than $\gamma_b= 10$. This indicates that a higher value of artificial noise strength is required to adequately capture the sudden and significant change in stiffness. Model $\mathcal{M}_5$ is has the highest probability among the remaining models ($79.71\%$) when the initial spring stiffness is known, followed by model $\mathcal{M}_6$ with $15.36\%$, highlighting the benefit of estimating the strength of the artificial noise by MCMC.

Finally, we consider a third row of model probabilities (denoted by $**$ in Table \ref{table:case2evidence}), where we also omit the models calibrated with a correct initial estimate for the system stiffness. Given that one of the objectives is to estimate the stiffness before and after the structural damage is incurred, it would seem logical that when considering the stiffness within the augmented state, that its initial mean may also be unknown. In model $\mathcal{M}_6$, we estimate the initial mean for the stiffness as a static parameter by MCMC, which provides superior estimates compared to models $\mathcal{M}_4^a$, $\mathcal{M}_4^b$ and $\mathcal{M}_5$, when the initial condition for the stiffness is incorrect. 

\begin{table}[!h]
{\footnotesize
\centering
\begin{tabular}{lrrrrrrrrrr}
\hline
 \Big. Model & $\mathcal{M}_1$ & $\mathcal{M}_2$ & $\mathcal{M}_3$ & \multicolumn{2}{c}{${\mathcal{M}_4}^a$} &  \multicolumn{2}{c}{${\mathcal{M}_4}^b$} & \multicolumn{2}{c}{$\mathcal{M}_5$} & $\mathcal{M}_6$ \\
\hline
\Big. log Evidence (-1600) 	&	56.16	&-104.07&	-652.78	&-138.62	&-109.94	&-11.82	&-14.54	&-3.06&	-5.84	&-4.71
 \\
 \Big. Average data-fit (-1600) 	&  81.19&-97.47&	-640.00	&-136.69&	-107.89	&-7.38	&-10.11&	8.05	&5.03&	7.54
\\
 \Big. Expected information gain &25.02	&6.60	&12.78	&1.94	&2.04	&4.45	&4.42	&11.11	&10.87&	12.25
\\
\hline
 \Big. Model probability (\%) &	100.00 &0.00 &0.00 &0.00 &	(0.00) &	0.00 &	(0.00) &	0.00&	(0.00)&	0.00 \\	
 \Big. Model probability$*$ (\%) &	-  &0.00 &0.00 &0.00 &	(0.00) &	0.01 &	(0.00) &	79.71&	(4.92)&	15.36 \\	
 \Big. Model probability${**}$ (\%) &	-  &0.00 &0.00 & - &	(0.00) &	- &	(0.00) &	 -&	(24.28)&		75.72\\	
\hline
\hline									
\end{tabular}
\caption{Model selection results for Case 2. Bracketed values correspond to poor estimates of the initial stiffness. The second row (*) of model probabilities summarizes the results if the generating model $\mathcal{M}_1$ is removed from the candidate set. The third row (**) of model probabilities summarizes the results where the models featuring correct initial means for the stiffness in models $\mathcal{M}_4^a$, $\mathcal{M}_4^b$ and $\mathcal{M}_5$ are also removed from the candidate set.}\label{table:case2evidence}
}
\end{table}

\subsection{Case 3 (Gradual change in stiffness)}\label{slowsnap}
\begin{figure}[H]
\centering
\includegraphics[width=\linewidth,keepaspectratio]{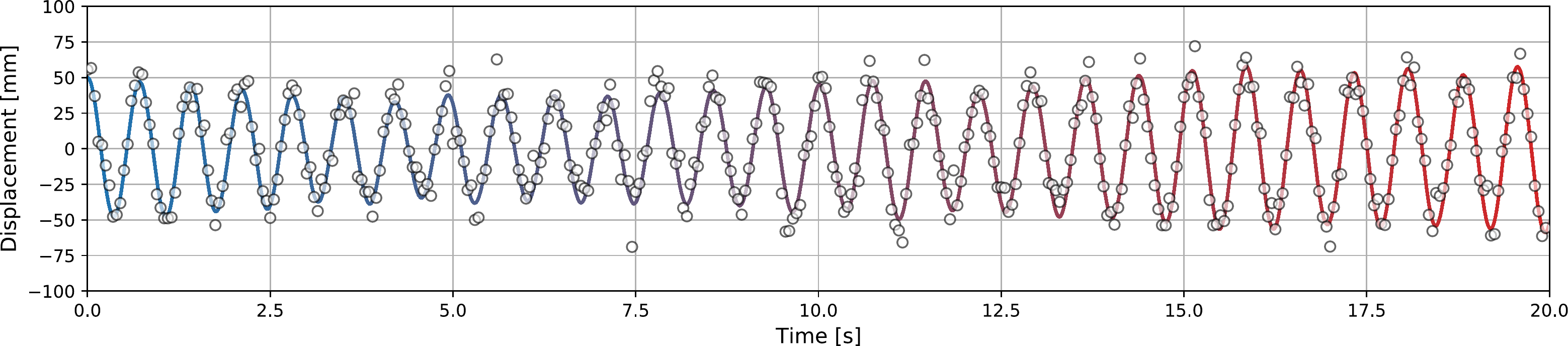}
\caption{Synthetic data for Case 3.}
\label{fig:case3}
\end{figure}

In this section, we highlight the lack of generality in the generating model for cases where the spring does not snap, but instead, degrades linearly over the period of observation. We consider the magnitude of degradation to be consistent with Case 1, where the initial system stiffness is 80 N/mm, and degrades to a value of 70 N/mm over the course of 20 seconds. Figure \ref{fig:case3} shows a realization of the system, and the synthetic data points that are generated from it. The colour gradient in the true signal signifies the change in stiffness from the undamaged state in blue (having a stiffness of 80 N/mm), to the damaged state in red (having a stiffness of 70 N/mm). 

\begin{figure}[H]
\centering
\includegraphics[width=\linewidth,keepaspectratio]{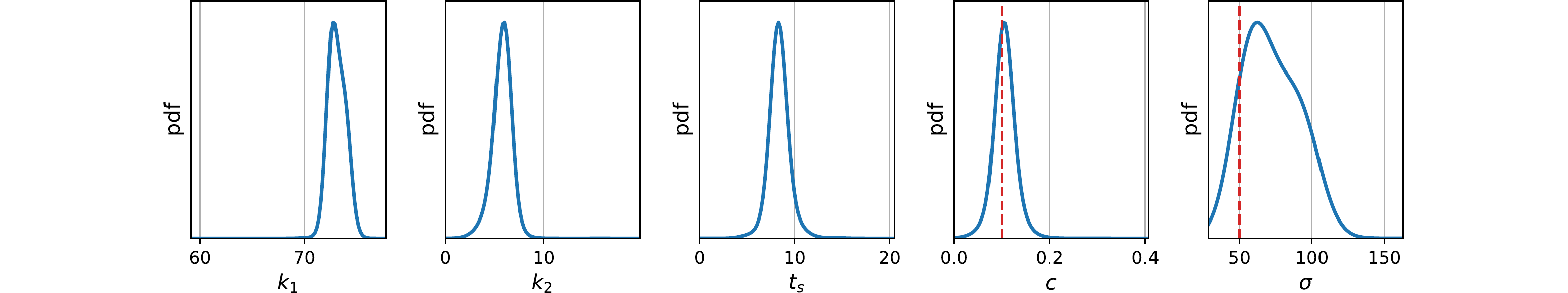}
\caption{Parameter posterior estimates of $\mathcal{M}_1$}\label{fig:c3m1pdf}
\end{figure}

Model $\mathcal{M}_1$ was designed to capture a sudden change in stiffness, however for this case where the stiffness decreases linearly over the entire period of observation. \textcolor{oct15}{The posterior estimates for the stiffness parameter are shown in Figure \ref{fig:c3m1pdf}. These estimates are understandably quite poor as the explicit model for the time-varying stiffness does not reflect the nature of the degradation of the springs.}

\begin{figure}[H]
\centering
\includegraphics[width=\linewidth,keepaspectratio]{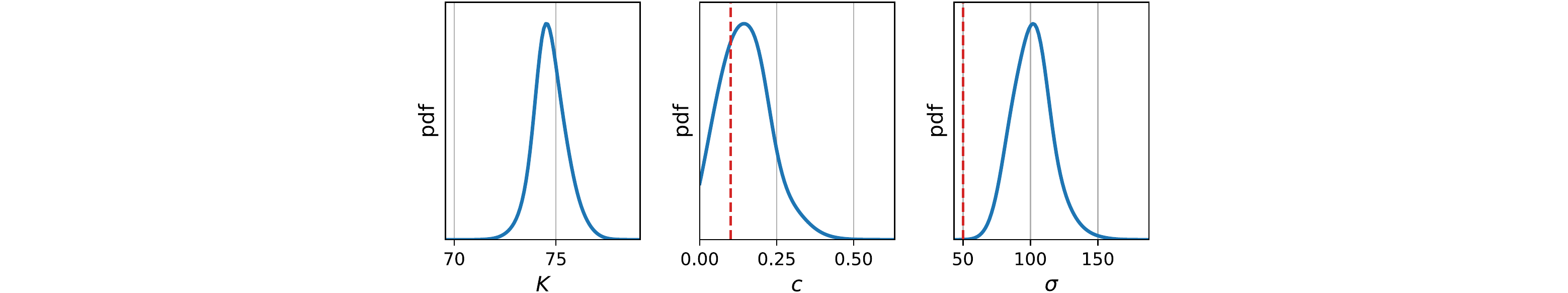}
\caption{Parameter posterior estimates of $\mathcal{M}_2$}\label{fig:c3m2pdf}
\end{figure}

\begin{figure}[H]
\centering
\includegraphics[width=\linewidth,keepaspectratio]{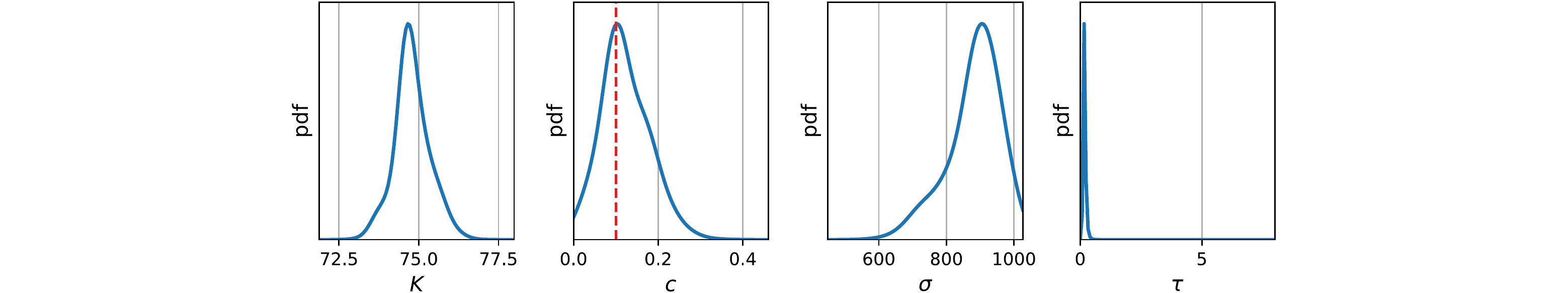}
\caption{Parameter posterior estimates of $\mathcal{M}_3$}\label{fig:c3m3pdf}
\end{figure}

 \textcolor{oct15}{In Figure \ref{fig:c3m2pdf}, the mode of the posterior distribution of the static stiffness parameter is approximately located at the average value of the stiffness over the 20 seconds of observation. Likewise, the model error strength is much higher than the stochastic forcing strength. The coloured noise model in $\mathcal{M}_3$ performs similarly to the white noise model in $\mathcal{M}_2$ as observed in Figure \ref{fig:c3m3pdf}}

\begin{figure}[H]
\centering
\begin{subfigure}{0.475\linewidth}
\includegraphics[width=\linewidth,keepaspectratio]{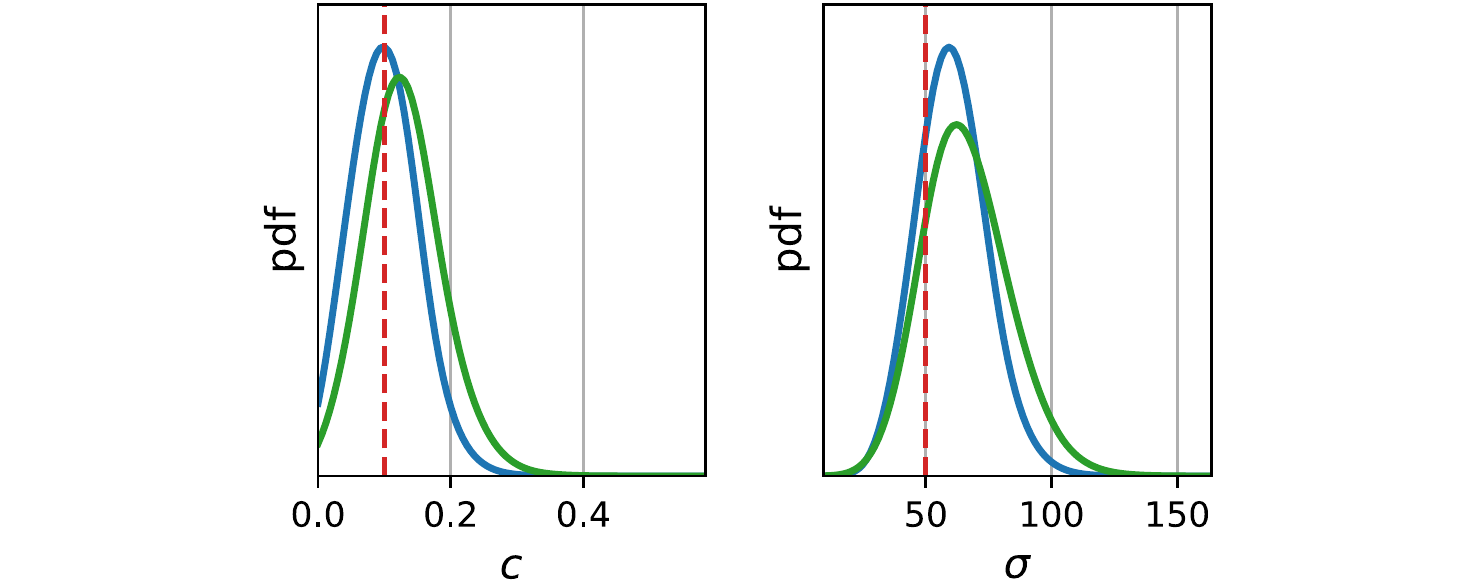}
\caption{Static parameter posterior estimates}\label{fig:c3m4apdf}
\end{subfigure}
\begin{subfigure}{0.475\linewidth}
\includegraphics[width=\linewidth,keepaspectratio]{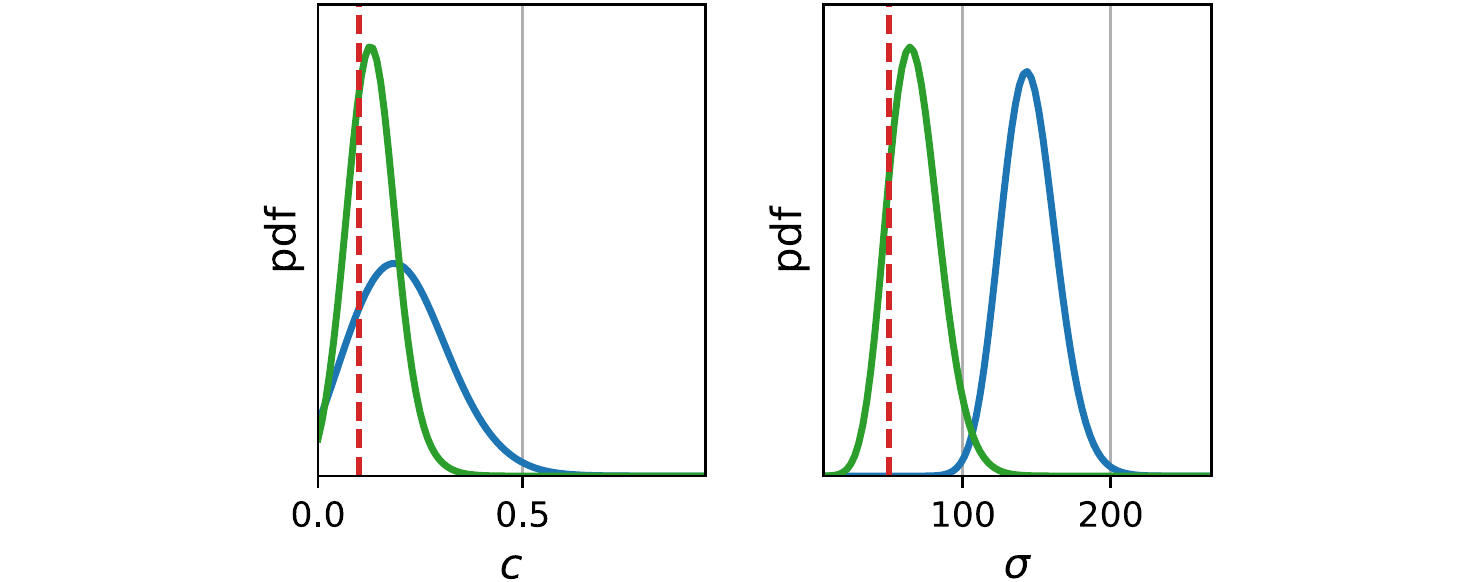}
\caption{Static parameter posterior estimates}\label{fig:c3m4bpdf}
\end{subfigure}
\begin{subfigure}{\linewidth}
\centering
\includegraphics[width=\linewidth,keepaspectratio]{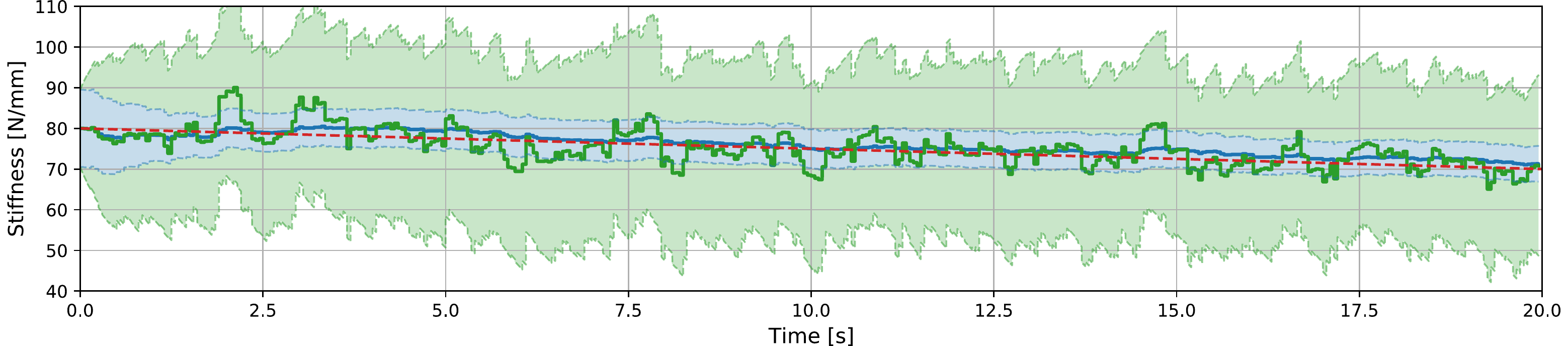}
\caption{EKF estimated mean (solid lines) $\pm$ 3 standard deviations (shaded area) for $K(t)$ at MAP estimates of static parameters}\label{fig:c3m4ase}
\end{subfigure}
\begin{subfigure}{\linewidth}
\centering
\includegraphics[width=\linewidth,keepaspectratio]{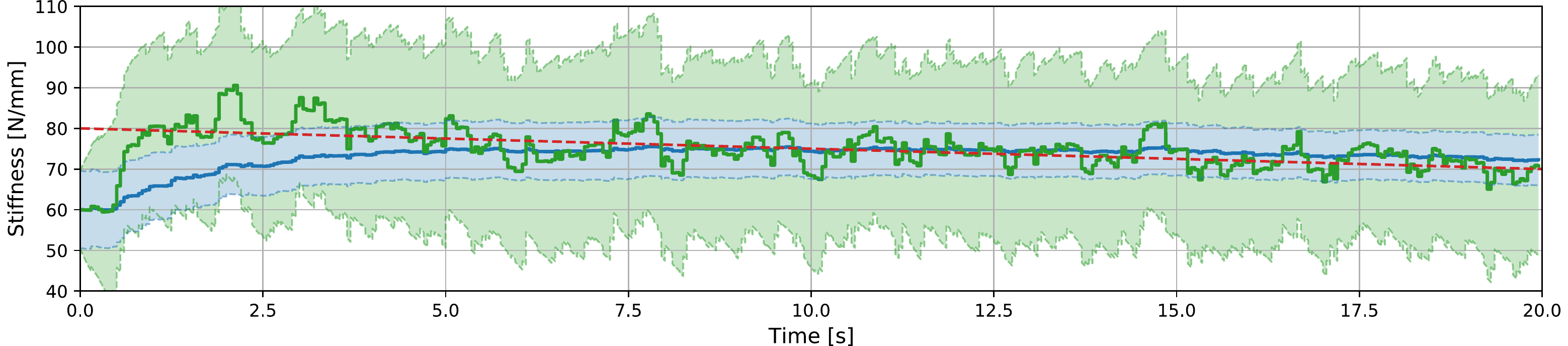}
\caption{EKF estimated mean (solid lines) $\pm$ 3 standard deviations (shaded area) for $K(t)$ at MAP estimates of static parameters}\label{fig:c3m4bse}
\end{subfigure}
\caption{Parameter estimates of $\mathcal{M}_4^a$ (blue) and $\mathcal{M}_4^b$ (green), and true parameter values (red), with initial mean $\mathbb{E}[K(0)] = 80$ (panels (\ref{fig:c3m4apdf}) and (\ref{fig:c3m4ase})), and with initial mean $\mathbb{E}[K(0)] = 60$ (panels (\ref{fig:c3m4bpdf}) and (\ref{fig:c3m4bse}))}\label{fig:c3m4all}
\end{figure}

For models $\mathcal{M}_4^a$ and $\mathcal{M}_4^b$, Figure \ref{fig:c3m4all} shows the results for two values of  the initial mean stiffness:
$\mathbb{E}[K(0)] = 80$  (correct value) and $\mathbb{E}[K(0)] = 60$ (incorrect value).   
The posterior pdfs of $c$ and $\sigma$ are very similar (in subplot (a)) for both models for $\mathbb{E}[K(0)] = 80$. However the additional uncertainty around the estimates (in subplot (c)) lowers the model probability for $\mathcal{M}_4^b$. Note from the subplot (d), the higher artificial noise strength in model $\mathcal{M}_4^b$ makes it more versatile, and allows it to quickly adapt to the incorrect initial condition ($\mathbb{E}[K(0)] = 60$).

\begin{figure}[H]
\centering
\begin{subfigure}{0.475\linewidth}
\includegraphics[width=\linewidth,keepaspectratio]{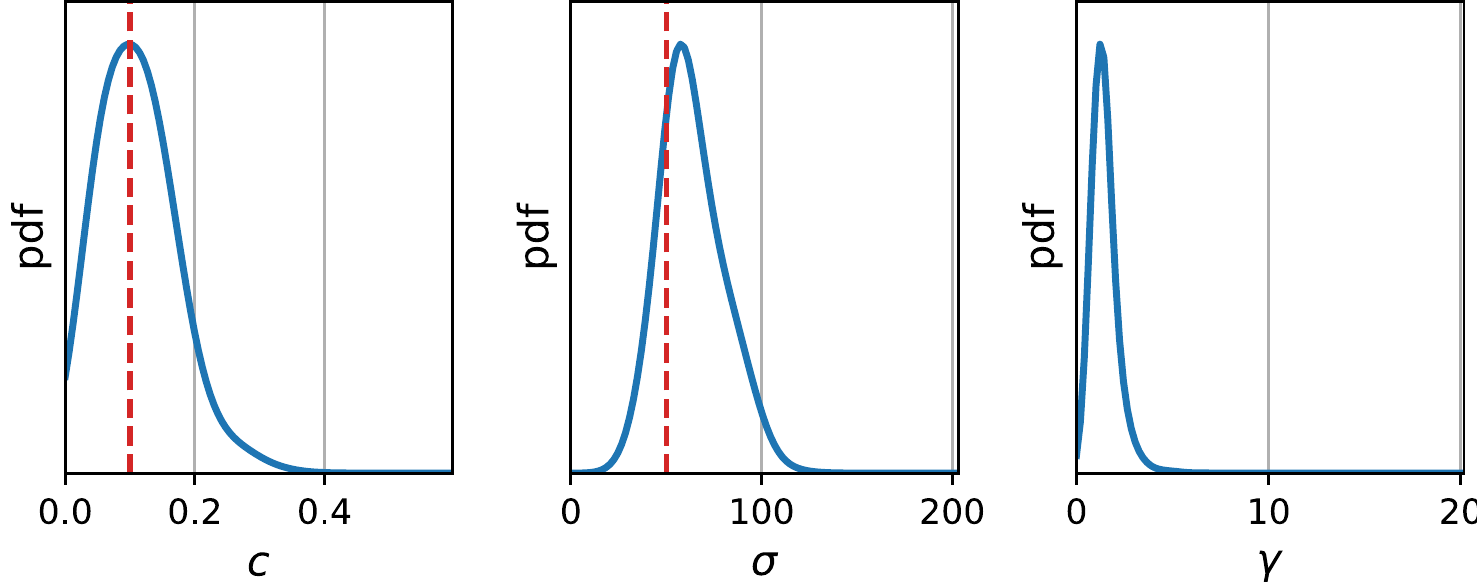}
\caption{Static parameter posterior estimates}\label{fig:c3m5apdf}
\end{subfigure}
\begin{subfigure}{0.475\linewidth}
\includegraphics[width=\linewidth,keepaspectratio]{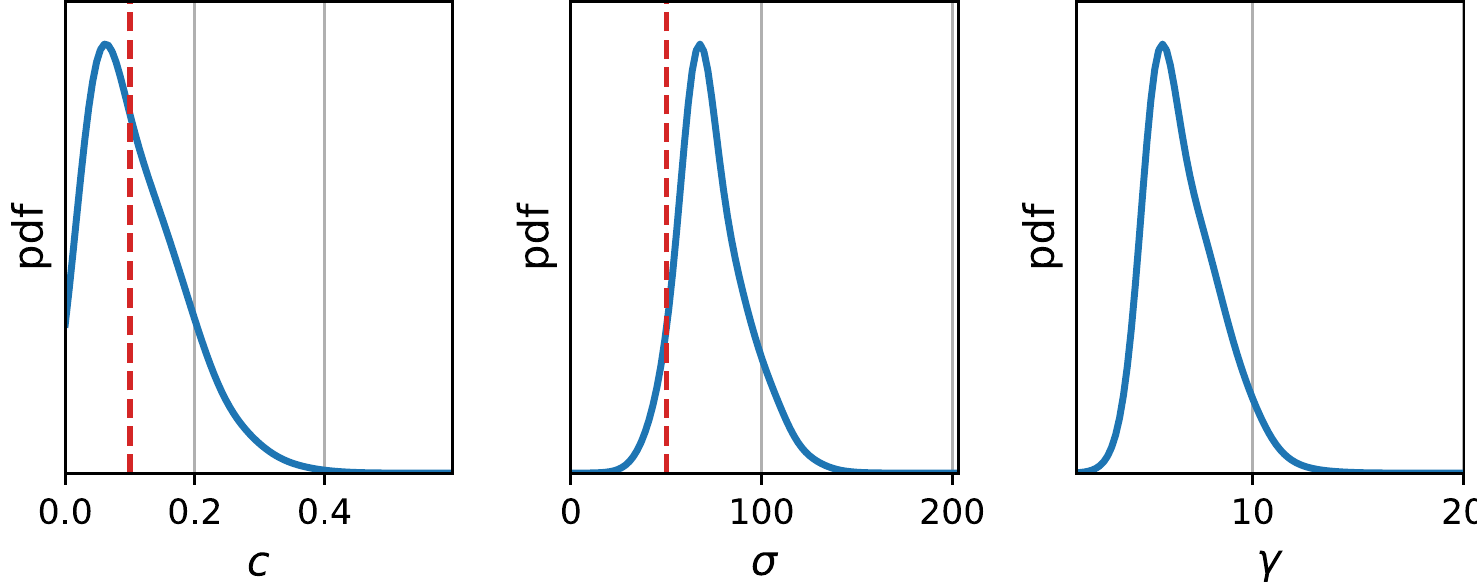}
\caption{Static parameter posterior estimates}\label{fig:c3m5bpdf}
\end{subfigure}
\begin{subfigure}{\linewidth}
\centering
\includegraphics[width=\linewidth,keepaspectratio]{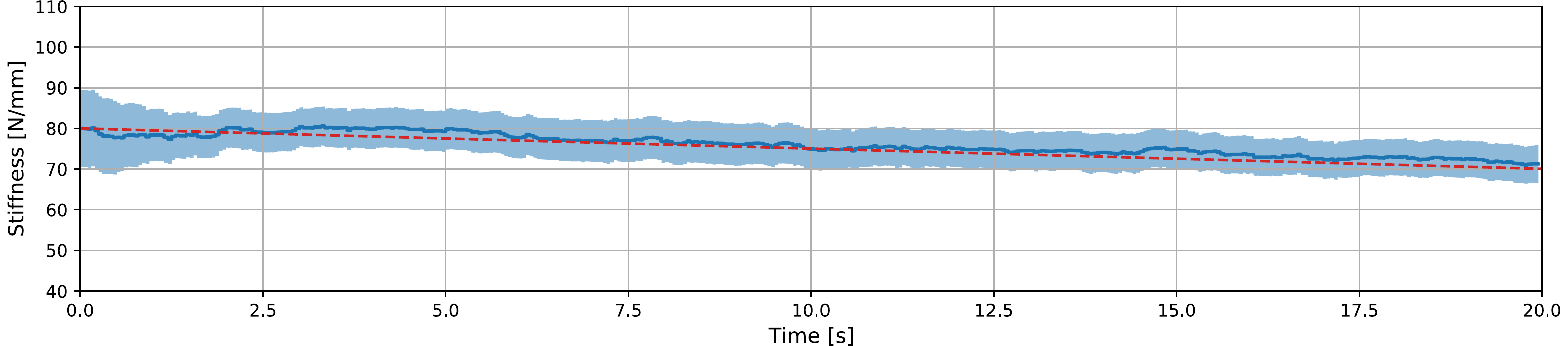}
\caption{EKF estimated mean (solid lines) $\pm$ 3 standard deviations (shaded area) for $K(t)$ at MAP estimates of static parameters}\label{fig:c3m5ase}
\end{subfigure}
\begin{subfigure}{\linewidth}
\centering
\includegraphics[width=\linewidth,keepaspectratio]{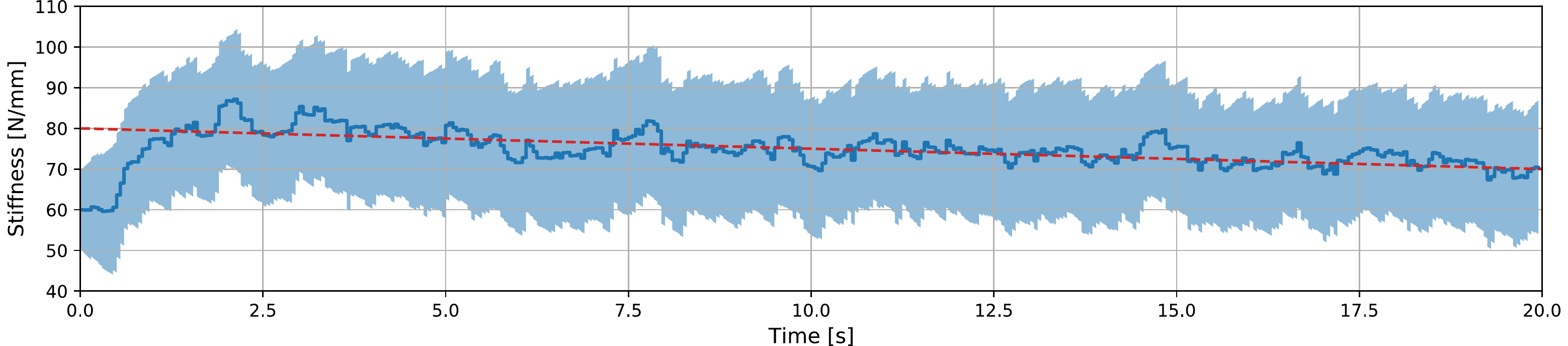}
\caption{EKF estimated mean (solid lines) $\pm$ 3 standard deviations (shaded area) for $K(t)$ at MAP estimates of static parameters}\label{fig:c3m5bse}
\end{subfigure}
\caption{Parameter estimates of $\mathcal{M}_5$ (blue), and true parameter values (red), with initial mean $\mathbb{E}[K(0)] = 80$ (panels (\ref{fig:c3m5apdf}) and (\ref{fig:c3m5ase})), and with initial mean $\mathbb{E}[K(0)] = 60$ (panels (\ref{fig:c3m5bpdf}) and (\ref{fig:c3m5bse}))}\label{fig:c3m5all}
\end{figure}

\begin{figure}[H]
\centering\centering
\begin{subfigure}{\linewidth}
\includegraphics[width=\linewidth,keepaspectratio]{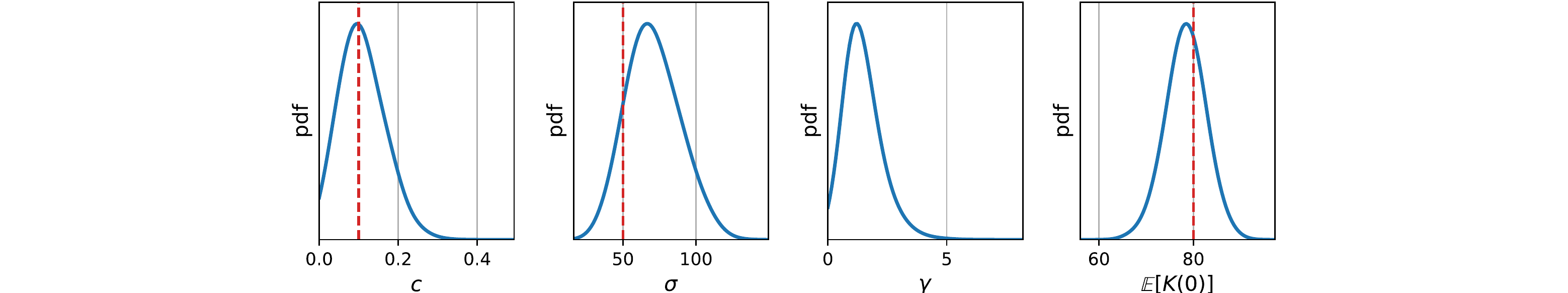}
\caption{Static parameter posterior estimates of $\mathcal{M}_6$}\label{fig:c3m6pdf}
\end{subfigure}
\begin{subfigure}{\linewidth}
\centering
\includegraphics[width=\linewidth,keepaspectratio]{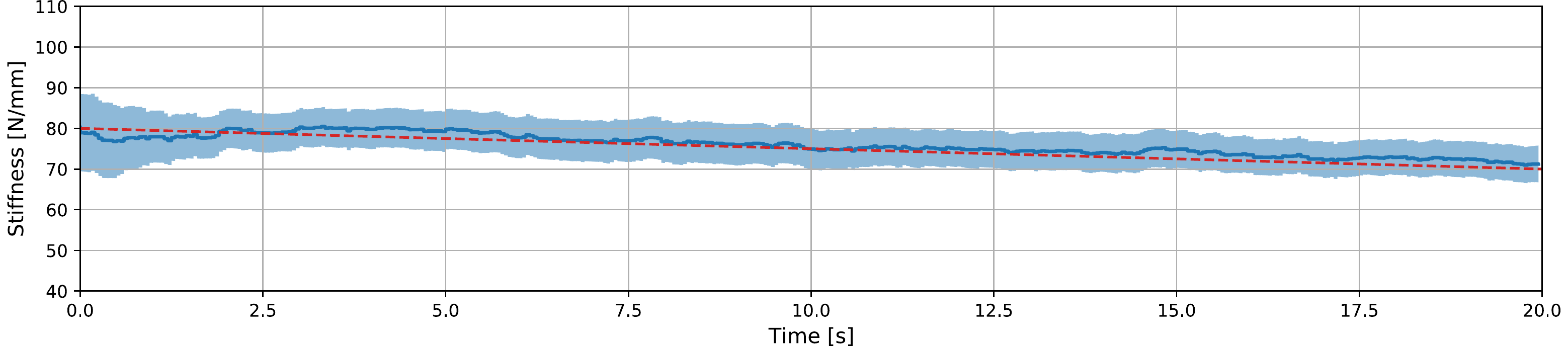}
\caption{EKF estimated mean (solid lines) $\pm$ 3 standard deviations (shaded area) for $K(t)$ at MAP estimates of static parameters}\label{fig:c3m6se}
\end{subfigure}
\caption{Parameter estimates of $\mathcal{M}_6$ (blue) and true parameter values (red)}\label{fig:c3m6all}
\end{figure}

Again for the two values of the initial mean stiffness:
$\mathbb{E}[K(0)] = 80$  and $\mathbb{E}[K(0)] = 60$,  when the artificial noise strength is estimated by MCMC in the model $\mathcal{M}_5$, the same trends are observed in Figure \ref{fig:c3m5all} compared to Figure \ref{fig:c3m4all}. When the initial mean stiffness is correct, the optimal noise strength is low. Conversely, when the initial the initial mean stiffness   is incorrect, a larger noise strength is preferred to quickly adjust to the correct value. 
Finally in the most general case of the estimation framework, when both the initial mean stiffness and artificial noise strength are estimated by MCMC, the estimate of the artificial noise strength is dictated by the  time-varying nature of the stiffness (rather than arbitrarily chosen values of the initial mean stiffsess) as demonstrated in Figure \ref{fig:c3m6all}.

\subsubsection{Model selection results and discussion}\label{c3_modelsel}
Interestingly, though model $\mathcal{M}_1$ no longer has the same form as the generating model, modelling a sudden change in stiffness rather than a gradual one, still has the best average data-fit among the candidate models. However, it is penalized heavily for its relatively high complexity. Ranked by data-fit, this model is followed closely by $\mathcal{M}_4^a$, $\mathcal{M}_5$ and $\mathcal{M}_6$, in descending order. Model $\mathcal{M}_4^a$ has the highest model probability as it is the simplest of these three models that estimate the stiffness as a time-varying parameter, using EKF. If we forego the assumption that the initial stiffness is well known (referring to the second row of model probabilities in Table \ref{table:case3evidence}, denoted by $*$), we see that model $\mathcal{M}_6$, has the highest probability.

\begin{table}[!h]
{\footnotesize
\centering
\begin{tabular}{lrrrrrrrrrr}
\hline
 \Big. Model & $\mathcal{M}_1$ & $\mathcal{M}_2$ & $\mathcal{M}_3$ & \multicolumn{2}{c}{${\mathcal{M}_4}^a$} &  \multicolumn{2}{c}{${\mathcal{M}_4}^b$} & \multicolumn{2}{c}{$\mathcal{M}_5$} & $\mathcal{M}_6$ \\
\hline
\Big. log Evidence (-1600) 	&	75.23	&73.85	&69.66	&86.47	&(70.78)	&70.78	&(68.68)	&79.69	&(67.61)	&77.33
 \\
 \Big. Average data-fit (-1600) 	&  93.69	&85.58	&84.08	&92.73	&(76.61)	&76.69	&(74.56)	&92.11	&(78.56)	&91.61
\\
 \Big. Expected information gain &18.46	&11.73	&14.43	& 6.25	& (5.83)	&5.90	&(5.89)	&12.42	&(10.95)&14.28
\\
\hline
 \Big. Model probability (\%) &	0.00 &0.00 &0.00 &	99.87 &	(0.00) &	0.00 &	(0.00) &	0.11 &	(0.00)&	0.01 \\	
  \Big. Model probability${*}$ (\%) &	10.58  & 2.66 &0.04 & - &	(0.12) &	- &	(0.02) &	 -&	(0.01)&		86.58\\	\hline
\hline									
\end{tabular}
\caption{Model selection results for Case 3. Bracketed values correspond to poor estimates of the initial stiffness. The second row (*) of model probabilities summarizes the results where the models featuring correct initial means for the stiffness in models $\mathcal{M}_4^a$, $\mathcal{M}_4^b$ and $\mathcal{M}_5$ are also removed from the candidate set.}\label{table:case3evidence}
}
\end{table}

\section{Conclusion}
We have presented a Bayesian model calibration framework leveraging both MCMC and nonlinear filters for the combined state and parameter estimation for dynamical systems containing both time-varying and time-invariant parameters. We generate synthetic data for three cases and demonstrate the proposed framework's performance in each situation: (i) for a sudden, subtle change in stiffness, where the effect on the dynamics is reasonably obscured by the noise in the measurements, (ii) for a sudden, significant change in stiffness, where the effects on the dynamics are visually evident, and (iii) for a gradual, subtle change in stiffness, where no model in the candidate set has the same functional form as the generating model.  We use a set of six candidate models, each having unique features to assess certain qualities of the proposed framework. We consider a baseline model ($\mathcal{M}_1$) which has the same analytical form as the generating model (in all cases but one), a model with only static parameters and a white noise error model ($\mathcal{M}_2$), a model with only static parameters and a coloured noise error model ($\mathcal{M}_3$), and three models leveraging the use of nonlinear filters to estimate time-varying parameters, while using MCMC to estimate the remaining time-invariant model parameters. Through the introduction of model $\mathcal{M}_4$, we have illustrated the benefit of capturing the evolution of time-varying parameters in time by concurrent state and parameter estimation within the MCMC framework, when the precise nature of time-varying parameters are not known (as in a sudden versus gradual decrease in stiffness). Subsequently, by the introduction of model $\mathcal{M}_5$, we have illustrated the benefit of the MCMC framework, for estimating the artificial noise strength driving the dynamics of the augmented state, for cases where the range of the time-varying parameter is unknown a priori (comparing results from Case 1 and Case 2). Finally, in model $\mathcal{M}_6$, we have illustrated the benefit of estimating the initial mean of the stiffness parameter in the augmented state as a time-invariant parameter. This could be of great benefit in a scenario where some unknown structural degradation may have already occurred, as models $\mathcal{M}_4$ and $\mathcal{M}_5$ under-performed compared to $\mathcal{M}_6$ when they assumed an incorrect initial stiffness.

\bibliographystyle{unsrt}
\bibliography{refs}{}

\end{document}